\documentclass[aps,prd,twocolumn,preprintnumbers,
superscriptaddress,showpacs,floatfix,nofootinbib]{revtex4}

\usepackage{bm}
\usepackage{epsfig}
\usepackage{graphics}

\usepackage{graphicx}

\begin{document}

\newcommand{\lsim}{\mbox{\raisebox{-.9ex}{~$\stackrel{\mbox{$<$}}{\sim}$~}}}

\newcommand{\gsim}{\mbox{\raisebox{-.9ex}{~$\stackrel{\mbox{$>$}}{\sim}$~}}}

\title{Low scale inflation and the curvaton mechanism}

\author{Konstantinos Dimopoulos}
\email{k.dimopoulos1@lancaster.ac.uk}
\affiliation{Department of Physics, Lancaster University, Lancaster LA1 4YB,
 UK}
\author{David H. Lyth}
\email{d.lyth@lancaster.ac.uk}
\affiliation{Department of Physics, Lancaster University, Lancaster LA1 4YB,
 UK}
\author{Yeinzon Rodr\'{\i}guez}
\email{y.rodriguezgarcia@lancaster.ac.uk}
\affiliation{Department of Physics, Lancaster University, Lancaster LA1 4YB,
 UK}
\affiliation{Centro de Investigaciones, Universidad Antonio Nari\~no, Cll 58A
 \# 37-94, Bogot\'a D.C., Colombia}

\date{\today}

\begin{abstract} 
The primordial curvature perturbation may be due to a `curvaton' field, 
which dominates (or almost dominates) the energy density
before it decays. In the simplest version of the curvaton model 
the scale of inflation has to be quite high corresponding to a 
Hubble parameter $H>10^7$\,GeV. We here explore
two  modifications of the curvaton model which can instead allow
inflation at a low scale. (i) The curvaton is a  Pseudo Nambu-Goldstone Boson 
(PNGB), with a  symmetry-breaking phase transition during inflation. (ii)
The curvaton mass increases suddenly at some moment after the end of inflation 
but before the onset of the curvaton oscillations. Both proposals can work but 
not in a completely natural way. Also, the lower bound on the scale of
inflation depends somewhat on the details of the framework used. Nevertheless,
we show that inflation with $H$ as low as 1~TeV or lower is possible to
attain.
\end{abstract}

\maketitle

\section{Introduction}

The primordial curvature perturbation is generated presumably from the 
perturbation of some scalar field, which in turn is generated from the 
vacuum fluctuation during inflation. The scalar field responsible for the
primordial curvature perturbation is traditionally supposed to be the 
inflaton field, i.e. the field responsible for the dynamics and, in particular,
the end of inflation \cite{book}. 
This `inflaton hypothesis' is economical, but it is 
quite difficult to implement and, if many scalar fields exist, it presumably 
is not particularly likely. An alternative is that the curvature perturbation 
is generated by a `curvaton' field, which dominates (or almost dominates) 
the energy density before it decays \cite{lyth02,moroi01} (see also
\cite{earlier}). 
According to this `curvaton hypothesis', the contribution of the inflaton
to the curvature perturbation is negligible. This is especially true if 
the energy scale of inflation is much lower than the scale of grand 
unification, which is the typical requirement of the traditional inflaton 
hypothesis\footnote{Although some exceptions exist, e.g. see Ref. \cite{snat}.}.
In fact, one of the advantages of the curvaton scenario is the relaxation
of the constraints on the inflationary energy scale, which alleviates many 
tuning problems in inflation model--building and allows for the construction 
of realistic, theoretically well-motivated inflation models~\cite{liber}.

In the simplest version of the curvaton model though,
the scale of inflation is still required to be quite high corresponding to 
Hubble parameter \mbox{$H>10^7$\,GeV} \cite{lyth04}. 
The purpose of this paper is to systematically
explore some modifications of the curvaton model which can instead allow
inflation at an even lower scale. To be specific, we aim for 
 $H\sim 10^3$\,GeV, which holds if the inflationary potential is 
generated by some mechanism of  gravity-mediated supersymmetry 
breaking which holds also in the vacuum.

We begin by presenting some known bounds in a unified notation.
Then we consider two possibilities: (i) that the curvaton is a
Pseudo Nambu-Goldstone Boson (PNGB), whose order parameter is increased 
after the cosmological scales exit the horizon during inflation but before the 
onset of the curvaton oscillations and (ii) that the  curvaton mass increases 
suddenly at some moment after the end of inflation but before the onset of 
the curvaton oscillations. We conclude with a prognosis for the viability
of low-scale inflation within the curvaton model.

Throughout our paper we use units such that \mbox{$\hbar=c=k_B=1$} and 
Newton's gravitational constant is \mbox{$8\pi G=m_P^{-2}$}, with 
\mbox{$m_P=2.4\times 10^{18}$ GeV} being the reduced Planck mass.

\section{The bounds on the scale of inflation}

In this section we present four bounds on the scale of inflation, in terms
of three parameters which encode possible modifications of the simplest 
curvaton scenario. These bounds have been presented at least implicitly in
earlier works \cite{lyth04,matsuda03,postma04} but not in the unified notation 
that we 
employ. The advantage of this notation is that it will allow us to compare the 
bounds in various situations, establishing with ease which is  the most 
crucial. The three parameters are
\begin{itemize}
\item
The  ratio $\varepsilon\equiv\sigma_*/\sigma_{\rm osc}$, where
$\sigma_*$ is the value of the curvaton field at horizon exit and 
$\sigma_{\rm osc}$ is its value when it starts to oscillate.
\item
The ratio $f\equiv H_{\rm osc}/\tilde m_\sigma$, where $H_{\rm osc}$ is the 
Hubble parameter at the start of the oscillation and $\tilde m_\sigma$
is the effective curvaton mass after the onset of the oscillation.
\item
The ratio $\delta\equiv\sqrt{H_{\rm osc}/H_*}$ where $H_*$ is the Hubble 
parameter during inflation.
\end{itemize}

\subsection{Curvaton physics considerations}

The  observed value of the nearly scale-invariant spectrum of curvature 
perturbations is 
${\cal P}_\zeta=(5\times 10^{-5})^2$, which we denote simply by $\zeta^2$.
In the curvaton scenario $\zeta$ is given by \cite{lyth02,lyth03c}
\begin{equation}
\zeta\sim\Omega_{\rm dec}\zeta_\sigma\;,
\label{zeta}
\end{equation}
where \mbox{$\Omega_{\rm dec}\leq 1$} is the density fraction of the curvaton
density $\rho_\sigma$ over the total density of the Universe $\rho$ at the time
of the decay of the curvaton:
\begin{equation}
\Omega_{\rm dec}\equiv\left.\frac{\rho_\sigma}{\rho}\right|_{\rm dec},
\label{r}
\end{equation}
and $\zeta_\sigma$ is the curvature perturbation of the curvaton field
$\sigma$, which is \cite{CD}
\begin{equation}
\zeta_\sigma\sim\left.\frac{\delta\sigma}{\sigma}\right|_{\rm dec}\sim
\left.\frac{\delta\sigma}{\sigma}\right|_{\rm osc}\;,
\label{zs1}
\end{equation}
where `osc' denotes the time when the curvaton oscillations begin and `dec'
denotes the time of curvaton decay.

In all the cases, which we consider,
\begin{equation}
\left.\frac{\delta\sigma}{\sigma}\right|_*\simeq
\left.\frac{\delta\sigma}{\sigma}\right|_{\rm osc}\,,
\label{ds/s}
\end{equation}
where `*' denotes the epoch when the cosmological scales exit the horizon 
during inflation. The above typically holds true because the curvaton (being a 
light field) is frozen during and after inflation until the onset of its 
oscillations. However, this does not mean that 
\mbox{$\sigma_*\simeq\sigma_{\rm osc}$} 
necessarily. Indeed, in the case of the PNGB curvaton (case (i)), with a 
varying order parameter $v$, the curvaton field is associated with the angular 
displacement $\theta$ from the minimum of its potential as 
\begin{equation}
\sigma\equiv\sqrt 2\,v\theta \;.
\label{theta}
\end{equation}
Therefore, even though after the end of inflation, $\theta$ remains 
approximately frozen (the angular motion is over damped), we may have 
\mbox{$\varepsilon\ll 1$}, where
\begin{equation}
\varepsilon\equiv\frac{\sigma_*}{\sigma_{\rm osc}}\,,
\label{eps}
\end{equation} 
because [cf. Eq.~(\ref{theta})] 
\mbox{$v_*=\varepsilon v_{\rm osc}\ll v_{\rm osc}$}. However, in this 
case too, for the curvaton fractional perturbation we find
\begin{equation}
\left.\frac{\delta\sigma}{\sigma}\right|_*
=\left.\frac{\delta\theta}{\theta}\,\right|_*\simeq
\left.\frac{\delta\sigma}{\sigma}\right|_{\rm osc},
\label{fractional}
\end{equation}
which agrees nicely with Eq.~(\ref{ds/s}). 

Now, for the perturbation of the curvaton we have
\begin{equation}
\delta\sigma_*=\frac{H_*}{2\pi}\,,
\label{dsH}
\end{equation}
Combining Eqs.~(\ref{eps}) and (\ref{dsH}) we find
\begin{equation}
\delta\sigma_{\rm osc}\simeq\frac{H_*}{2\pi\varepsilon}\,,
\label{dsosc}
\end{equation}
which means that, if the order parameter of a PNGB curvaton grows, the curvaton
perturbation is amplified by a factor $\varepsilon^{-1}$.

From Eqs.~(\ref{zeta}) and (\ref{zs1}) we have
\begin{equation}
\sigma_{\rm osc}\sim (\Omega_{\rm dec}/\zeta)\,\delta\sigma_{\rm osc}\;.
\end{equation}
Using Eq.~(\ref{dsosc}), we can recast the above as
\begin{equation}
\sigma_{\rm osc}\sim\frac{H_*\Omega_{\rm dec}}{\pi\varepsilon\zeta}\,.
\label{sosc}
\end{equation}

\subsection{The main bound on the scale of inflation}

For the density fraction at the onset of the curvaton oscillations we have:
\begin{equation}
\left.\frac{\rho_\sigma}{\rho}\right|_{\rm osc}\sim f^{-2}
\left(\frac{\sigma_{\rm osc}}{m_P}\right)^2,
\label{rhofracosc}
\end{equation}
where 
\begin{equation}
f\equiv\frac{H_{\rm osc}}{\tilde m_\sigma},
\label{f}
\end{equation}
and
we used that 
\mbox{$(\rho_\sigma)_{\rm osc}\simeq
\frac{1}{2}\tilde m_\sigma^2\sigma_{\rm osc}^2$} 
and \mbox{$\rho_{\rm osc}=3H_{\rm osc}^2m_P^2$}. Here, $\tilde m_\sigma$
denotes the effective mass of the curvaton {\em after} the onset of its 
oscillations. In the basic setup of the curvaton hypothesis this effective mass
 is the bare mass $m_\sigma$. If this is the case then 
\mbox{$\tilde m_\sigma=m_\sigma\simeq H_{\rm osc}$} (i.e. \mbox{$f\simeq 1$}).
However, in the heavy curvaton scenario (case (ii)), the mass of the curvaton
is supposed to be suddenly incremented at some time after the end of the 
inflationary epoch due to a coupling of the form 
$\lambda \chi^2 \sigma^2$ with a field $\chi$ which acquires a large vacuum
 expectation value (VEV) at some time after the end of inflation. In this case
\mbox{$\tilde{m}_\sigma^2=m_\sigma^2+\lambda\langle\chi\rangle^2\approx 
\lambda\langle\chi\rangle^2\gg H_{\rm osc}^2$} (i.e. \mbox{$f\ll 1$}).

Now, we need to consider separately the cases when the curvaton decays before
it dominates the Universe (\mbox{$\Omega_{\rm dec}\ll 1$}) or after it does
so (\mbox{$\Omega_{\rm dec}\sim 1$}). Note, that the WMAP constraints 
on non-gaussianity in the Cosmic Microwave Background
Radiation (CMBR) impose a lower bound on $\Omega_{\rm dec}$, which allows the
range \cite{lyth03c, komatsu03}:
\begin{equation}
0.01 \leq \Omega_{\rm{dec}} \leq 1. 
\label{WMAPr}
\end{equation} 
Because of the above bound we require that the density ratio $\rho_\sigma/\rho$
grows substantially after the end of inflation. Typically, in
the curvaton scenario this does indeed take place after the curvaton begins
oscillating, but only if the curvaton oscillates in a quadratic potential 
during the radiation era. As it was shown in Ref.~\cite{CD}, if the curvaton 
oscillates in a quartic or even higher order potential, its density ratio does 
not increase with time (it may well decrease instead) and satisfying the bound 
in Eq.~(\ref{WMAPr}) is very hard. Due to this fact, in the following, unless 
stated otherwise, we assume that at least a part of the period of oscillations 
occurs in the radiation era with a quadratic potential. Hence, we consider that
\mbox{$H_{\rm dec}\leq\Gamma_{\rm inf}$}. We also consider the minimal scenario
that the Universe, after the end of inflation, undergoes a period of matter
domination (due to coherent inflaton oscillations) until reheating, when it
becomes radiation dominated.

Suppose, at first, that the curvaton decays before dominating the density of
the Universe so that \mbox{$\Omega_{\rm dec}\ll 1$}.
Assuming that the curvaton oscillates in a quadratic potential, during the 
radiation epoch, its density fraction grows as 
\mbox{$\rho_\sigma/\rho\propto H^{-1/2}$}. 
Therefore, at curvaton decay we have
\begin{equation}
\Omega_{\rm{dec}}\sim\min\left\{1,\sqrt{\Gamma_{\rm inf}/H_{\rm osc}}\right\}
\frac{\tilde m_\sigma^2\sigma_{\rm{osc}}^2}{T_{\rm{dec}}
H_{\rm{osc}}^{3/2}m_P^{3/2}}\;, 
\label{Tdec}
\end{equation}
where we used Eq.~(\ref{rhofracosc}) and also that 
\mbox{$\rho_{\rm dec}\sim T_{\rm dec}^4$}, with
$\Gamma_{\rm inf}$ being the 
decay rate of the inflaton field, which determines the reheat temperature as
\mbox{$T_{\rm reh}\sim\sqrt{\Gamma_{\rm inf}\,m_P}$}. Using Eq.~(\ref{sosc})
the above can be recast as
\begin{equation}
H_*\sim\pi\varepsilon\zeta f
\frac{m_P}{\sqrt{\Omega_{\rm dec}}}
\left(\frac{H_{\rm dec}}{H_{\rm osc}}\right)^{1/4}
\max\left\{1,
\frac{H_{\rm osc}}{\Gamma_{\rm inf}}
\right\}^{1/4},
\label{H*1}
\end{equation}
where we used that \mbox{$T_{\rm dec}^2\sim H_{\rm dec}m_P$}.

Now, suppose that the curvaton decays after it dominates the Universe so that
\mbox{$\Omega_{\rm dec}\sim 1$}. Since 
\mbox{$(\rho_\sigma/\rho)_{\rm dom}\simeq 1$}
by definition, using again that, during the radiation epoch, 
\mbox{$\rho_\sigma/\rho\propto H^{-1/2}$} and in view of 
Eq.~(\ref{rhofracosc}), we obtain
\begin{equation}
H_{\rm dom}\sim\min\{H_{\rm osc}, \Gamma_{\rm inf}\}
f^{-4}
\left(\frac{\sigma_{\rm osc}}{m_P}\right)^4,
\label{Hdom}
\end{equation}
where `dom' denotes the time of curvaton domination. Employing again 
Eq.~(\ref{sosc}), the above can be written as
\begin{equation}
H_*\sim\pi\varepsilon\zeta
f m_P
\left(\frac{H_{\rm dom}}{H_{\rm osc}}\right)^{1/4}
\max\left\{1,
\frac{H_{\rm osc}}{\Gamma_{\rm inf}}
\right\}^{1/4}.
\label{H*2}
\end{equation}

Combining Eqs.~(\ref{H*1}) and (\ref{H*2}) we find that, in all cases
\begin{eqnarray}
H_* & \sim & \pi\varepsilon\zeta f\frac{m_P}{\sqrt{\Omega_{\rm dec}}}
\left(\frac{\max\{H_{\rm dom}, H_{\rm dec}\}}{H_{\rm osc}}\right)^{1/4}
\times\nonumber\\
 & \times & 
\max\left\{1,\frac{H_{\rm osc}}{\Gamma_{\rm inf}}\right\}^{1/4}.
\label{H*0}
\end{eqnarray}
This can be rewritten as 
\begin{eqnarray}
H_* & \sim & \Omega_{\rm dec}^{-2/5}
\left(\frac{H_*}{\min\{H_{\rm osc}, \Gamma_{\rm inf}\}}\right)^{1/5}
\times\nonumber\\
 &  & \hspace{-1.6cm}\times\;
\left(\frac{\max\{H_{\rm dom}, H_{\rm dec}\}}{H_{\rm BBN}}\right)^{1/5}
\!\!(\pi\varepsilon\zeta f)^{4/5}(T_{\rm BBN}^2m_P^3)^{1/5},
\label{H*}
\end{eqnarray}
or equivalently (using \mbox{$V_*^{1/4}\sim\sqrt{H_*m_P}$})
\begin{eqnarray}
V_*^{1/4} & \sim & \Omega_{\rm dec}^{-1/5}
\left(\frac{H_*}{\min\{H_{\rm osc}, \Gamma_{\rm inf}\}}\right)^{1/10}
\times\nonumber\\
 &  & \hspace{-2cm}\times\;
\left(\frac{\max\{H_{\rm dom}, H_{\rm dec}\}}{H_{\rm BBN}}\right)^{1/10}
\!\!(\pi\varepsilon\zeta f)^{2/5}(T_{\rm BBN}m_P^4)^{1/5},
\label{V*}
\end{eqnarray}
where `BBN' denotes the epoch of Big Bang Nucleosynthesis (BBN)
(\mbox{$T_{\rm BBN}\sim 1$ MeV}).
Now, according to Eq.~(\ref{WMAPr}) we have \mbox{$\Omega_{\rm dec}\leq 1$}.
Also, we require that the curvaton decays before BBN, i.e. 
\mbox{$H_{\rm dec}>H_{\rm BBN}$}. Moreover, we also 
have \mbox{$\Gamma_{\rm inf}\leq H_*$}. Hence,
the above provides the following bounds
\begin{equation}
\mbox{\framebox{%
\begin{tabular}{c}\\
$H_*>(\pi\varepsilon\zeta f)^{4/5}(T_{\rm BBN}^2m_P^3)^{1/5}\sim
(\varepsilon f)^{4/5}\times 10^7 \ {\rm GeV},$\\
\\
$V_*^{1/4}>(\pi\varepsilon\zeta f)^{2/5}(T_{\rm BBN}m_P^4)^{1/5}\sim
(\varepsilon f)^{2/5}\times 10^{12} \ {\rm GeV}$.\\
\\
\end{tabular}}}
\label{H*bound}
\label{V*bound}
\end{equation}
In the standard setup of the curvaton scenario \mbox{$\varepsilon=f=1$} and
the above bounds do not allow inflation at low energy scales to take place
\cite{lyth04}. However, we see that if either $\varepsilon$ or $f$ are much
smaller than unity the lower bound on the inflationary scale can be 
substantially relaxed and low scale inflation can be accommodated. Still, 
though, there are more bounds to be considered.

\subsection{Other bounds related to curvaton decay}

Firstly, let us consider the bound coming from the fact that the decay rate
of the curvaton field cannot be arbitrarily small. Indeed, in view of 
the fact that the curvaton interactions are at least of gravitational strength,
we find the following decay rate for the curvaton
\begin{equation}
\Gamma_\sigma \approx 
\gamma_\sigma 
\frac{\tilde{m}_\sigma^3}{m_P^2}\leq\tilde m_\sigma, 
\label{decay_rate}
\label{Gs}
\end{equation}
where \mbox{$\gamma_\sigma \gsim 1$}. 

Suppose, at first, that the curvaton decays after the onset of its 
oscillations, as in the basic setup of the curvaton scenario. In this case, 
\mbox{$\Gamma_\sigma\leq H_{\rm osc}$} and \mbox{$H_{\rm dec}=\Gamma_\sigma$}.
Using the fact that 
\mbox{max$\{H_{\rm dom}, \Gamma_\sigma\}\geq\Gamma_\sigma$},
Eq.~(\ref{decay_rate}) suggests
\begin{equation}
\frac{\max\{H_{\rm dom}, H_{\rm dec}\}}{H_{\rm osc}}\geq 
\gamma_\sigma 
f^{-1}
\left(\frac{\tilde m_\sigma}{m_P}\right)^2.
\end{equation}
Including the above into Eq.~(\ref{H*0}) the latter becomes
\begin{equation}
\,\!
H_*\!\geq\! 
\sqrt{\gamma_\sigma}(\pi\varepsilon\zeta)^2\sqrt{f}\,
\frac{m_P}{\Omega_{\rm dec}}\left(\frac{H_{\rm osc}}{H_*}\right)
\max\!\left\{\!1,
\frac{H_{\rm osc}}{\Gamma_{\rm inf}}
\right\}^{1/2}\hspace{-.5cm},\hspace{-.1cm}
\label{Hbound0}
\end{equation}
which results in the bounds
\begin{equation}
\mbox{\framebox{%
\begin{tabular}{c}\\
$H_*\geq 
(\pi\varepsilon\zeta)^2\sqrt{f}\;\delta^2\,m_P
\sim
\varepsilon^2\sqrt f\;\delta^2
\times 10^{11} \ {\rm GeV},$\\
\\
$V_*^{1/4}\geq\pi\varepsilon\zeta f^{1/4}\delta\,m_P
\sim
\varepsilon f^{1/4}\delta
\times 10^{14} \ {\rm GeV}$  ,\\
\\
\end{tabular}}}
\label{Hbound}
\label{Vbound}
\end{equation}
where we have defined 
\begin{equation}
\delta\equiv
\sqrt{\frac{H_{\rm{osc}}}{H_\ast}}\leq 1\;.
\label{dratio}
\label{delta}
\end{equation}
In the case of a PNGB curvaton (case (i)) we see that the bounds in 
Eq.~(\ref{Hbound}) are drastically reduced with $\varepsilon$, compared with
the bounds in Eq.~(\ref{H*bound}). This is not so in the heavy curvaton case
(case (ii)), where it is also possible that $\delta$ is not very small.

Now, provided we demand that the curvaton field does not itself result in a
period of inflation, we see that the curvaton cannot dominate the Universe 
before the onset of its oscillations. This results into the constraint
\begin{equation}
\,\!
\left.\frac{\rho_\sigma}{\rho}\right|_{\rm osc}\!\!\!\!\leq 1
\;\Leftrightarrow\;
\tilde m_\sigma\leq\pi\varepsilon\zeta\,\delta^2\frac{m_P}{\Omega_{\rm dec}}\,
\;\Leftrightarrow\;
f\geq\frac{\Omega_{\rm dec}H_*}{(\pi\varepsilon\zeta)m_P},\hspace{-.5cm}
\label{cons_m_n}
\label{mfbound}
\end{equation}
where we used Eqs.~(\ref{sosc}), (\ref{rhofracosc}), (\ref{f}) and 
(\ref{delta}). Inserting the above into Eq.~(\ref{Hbound0}) we obtain
\begin{equation}
H_*\geq
\gamma_\sigma(\pi\varepsilon\zeta)^3\delta^4\,
\frac{m_P}{\Omega_{\rm dec}}
\max\left\{1, H_{\rm osc}/\Gamma_{\rm inf}\right\},
\label{Hbound1}
\end{equation}
which results in the bounds
\begin{equation}
\mbox{\framebox{%
\begin{tabular}{c}\\
$H_*\geq (\pi\varepsilon\zeta)^3\delta^4m_P\sim
\varepsilon^3\delta^4\times 10^7 \ {\rm GeV},$\\
\\
$V_*^{1/4}\geq 
(\pi\varepsilon\zeta)^{3/2}\delta^2m_P\sim
\varepsilon^{3/2}\delta^2\times 10^{12} \ {\rm GeV}$ .\\
\\
\end{tabular}}}
\label{Hbound-0}
\label{first}
\label{Vbound-0}
\end{equation}
A similar bound is reached with the use of the upper bound on 
$\tilde{m}_\sigma$
\begin{equation}
\tilde m_\sigma\leq\gamma_\sigma^{-1/3}(H_{\rm osc}m_P^2)^{1/3},
\label{msbound}
\label{m_dr}
\end{equation}
which comes from $\Gamma_\sigma\leq H_{\rm osc}$ and the 
Eq.~(\ref{decay_rate}), 
instead of the bound in Eq. (\ref{cons_m_n}). Inserting the above into 
Eq.~(\ref{Hbound0}) one finds [cf. Eq.~(\ref{Hbound1})]
\begin{equation}
H_*\geq
\gamma_\sigma(\pi\varepsilon\zeta)^3\delta^4\,
\frac{m_P}{\Omega_{\rm dec}^{3/2}}
\max\left\{1,H_{\rm osc}/\Gamma_{\rm inf}\right\}^{3/4},
\label{Hbound2}
\end{equation}
which, again, results into the bound in Eq.~(\ref{Hbound-0}), as it was 
suggested in Ref.~\cite{postma04}.

In the heavy curvaton scenario (case (ii)) we have \mbox{$\varepsilon=1$} and
also \mbox{$H_{\rm osc}\simeq\min\{H_{\rm pt}, \tilde m_\sigma\}$}, where 
$H_{\rm pt}$ corresponds to the phase transition which increases the effective
mass of the curvaton. Then, if \mbox{$\delta\rightarrow 1$}, the bounds in 
Eq.~(\ref{Hbound-0})
are not possible to be relaxed below the standard case discussed in 
Ref.~\cite{lyth04} despite the fact that we may have \mbox{$f\ll 1$} in 
Eq.~(\ref{H*bound}). 
Therefore, in the heavy curvaton 
scenario we require \mbox{$\delta\ll 1$}, i.e. {\em the onset of the curvaton
oscillations has to occur much later than the end of inflation} so that 
\mbox{$H_*\gg H_{\rm osc}\geq\Gamma_\sigma$}. In this case, as can be seen in
Eq.~(\ref{Hbound-0}), 
it is easy to lower the bound
on the inflationary scale even for a not-so-small value of $\delta$.
This is a very nice feature of this scenario. Note also, that in the case of a 
PNGB curvaton (case (i)) \mbox{$H_{\rm osc}\sim m_\sigma\ll H_*$} and $\delta$
is very small necessarily. Because, in this case, \mbox{$\varepsilon\ll 1$},
it is straightforward to see that the bounds in Eq.~(\ref{Hbound-0}) are much
weaker than the bounds in Eq.~(\ref{H*bound}).

As it was pointed out in Ref.~\cite{postma04}, the sudden increment in the 
curvaton mass might lead to a growth in the curvaton decay rate enough for 
\mbox{$\Gamma_\sigma>H_{\rm pt}$}. This would force the curvaton to decay 
immediately and we can write 
\mbox{$H_{\rm osc}\sim H_{\rm pt}\sim H_{\rm dec}$}.
Obviously, in this case we cannot have \mbox{$H_{\rm dec}<H_{\rm dom}$} and
there is no period when \mbox{$\rho_\sigma/\rho\propto H^{-1/2}$}. This 
means that \mbox{$(\rho_\sigma/\rho)_{\rm osc}\sim\Omega_{\rm dec}$}. Using
Eqs.~(\ref{sosc}) and (\ref{rhofracosc}) it is easy to find
\begin{equation}
H_*\sim\pi\varepsilon\zeta 
f
\frac{m_P}{\sqrt{\Omega_{\rm dec}}},
\label{H*00}
\end{equation}
which results in the following bounds
\begin{equation}
\mbox{\framebox{%
\begin{tabular}{c}\\
$H_*\geq\pi\varepsilon\zeta f\,m_P\sim\varepsilon f\times 10^{14} 
\ {\rm GeV},$\\
\\
$V_*^{1/4}\geq\sqrt{\pi\varepsilon\zeta f}\,m_P\sim
(\varepsilon f)^{1/2}\times 10^{16} \ {\rm GeV} .$\\
\\
\end{tabular}}}
\label{Hbound-00}
\label{second}
\label{Vbound-00}
\end{equation}
It is evident that the above bounds may challenge the COBE constraint for the
curvaton scenario \cite{liber} and may lead to excessive curvature 
perturbations from the inflaton field if $\varepsilon$ and/or $f$ are not much 
smaller than unity. 

Note, that the bounds in Eq.~(\ref{Hbound-00}) are, in general, valid in the
case when \mbox{$H_{\rm dec}>\Gamma_{\rm inf}$} because, in this case
\mbox{$(\rho_\sigma/\rho)_{\rm osc}\sim\Omega_{\rm dec}$}. Indeed, combining
Eqs.~(\ref{H*0}) and (\ref{H*00}) we obtain the generic condition
\begin{eqnarray}
H_* & \sim & \pi\varepsilon\zeta f
\frac{m_P}{\sqrt{\Omega_{\rm dec}}}
\left(\frac{\max\{H_{\rm dec}, H_{\rm dom}\}}{H_{\rm osc}}\right)^{1/4}
\times\nonumber\\
 & & \times\;
\min\left\{1,\frac{\Gamma_{\rm inf}}{H_{\rm dec}}\right\}^{1/4}\!
\max\left\{1,\frac{H_{\rm osc}}{\Gamma_{\rm inf}}\right\}^{1/4}\!\!\!.
\label{H*000}
\end{eqnarray}

The bounds in Eqs. (\ref{H*bound}), (\ref{Hbound}), and  (\ref{H*00}) 
 provide the basis for our investigation, leaving the fourth bound in Eq.
(\ref{Hbound-00}) to be considered in a companion paper \cite{yeinzon04}.
As a matter of completeness we have considered all the other 
possible bounds coming from the requirements that 
\mbox{$\Gamma_\sigma < \tilde{m}_\sigma$} and 
\mbox{$H_{\rm dec} \geq H_{\rm BBN}$}. We have found that these bounds lead
to consistent and/or weaker constraints than the above four.

\section{The case of a PNGB curvaton with a varying order parameter}

We discuss here the case of a pseudo Nambu-Goldstone boson (PNGB) as curvaton.
Such a curvaton has the additional advantage of being protected by the
(approximate) global U(1) symmetry, which means that its flatness can be 
protected by the effect of supergravity corrections. Candidates for such a
PNGB curvaton have been discussed in Ref.~\cite{pngb}. However, in contrast to 
those models, we consider a PNGB curvaton, whose 
radial field has a larger expectation value in the vacuum than at the time when
the cosmological scales exit the horizon during inflation. Thus, the potential
for the curvaton field $\sigma$ is
\begin{eqnarray}
\hspace{-.5cm}
V(\sigma)=(v\tilde{m}_\sigma)^2\!\!
\left[1\!-\!\cos\left(\frac{\sigma}{v}\right)\right]
& \!\!\!\Rightarrow
V(|\sigma|\!<\!v)\simeq\frac{1}{2}\tilde{m}_\sigma^2\sigma^2\!, &
\label{Vs}
\end{eqnarray}
where \mbox{$v=v(t)$} is the expectation value of the radial field and
\mbox{$\tilde{m}_\sigma=\tilde{m}_\sigma(v)$} is the mass of the curvaton at
a given moment. In the true vacuum we have \mbox{$v=v_0$} and 
\mbox{$\tilde{m}_\sigma=m_\sigma$} with $v_0$ being the VEV of the radial field
and $m_\sigma$ being the mass of the curvaton in the vacuum.

To simplify our study {\em we assume that 
the curvaton mass has already assumed its vacuum value before the onset of 
the curvaton oscillations}. This means that at the onset of the curvaton 
oscillations \mbox{$v\rightarrow v_0$} and the mass of the curvaton has
its vacuum value $m_\sigma$. We further assume that \mbox{$m_\sigma\leq H_*$}
so that the oscillations begin after the end of inflation (were it otherwise 
the curvaton density would be exponentially diluted). Therefore, the curvaton 
oscillations begin when \mbox{$H_{\rm osc}\sim m_\sigma$}, which means that
\mbox{$f\sim 1$} and \mbox{$\delta\ll 1$} [cf. Eqs.~(\ref{f}) and (\ref{delta})
respectively]. Consequently, in view of Eqs.~(\ref{V*bound}), (\ref{Vbound})
and (\ref{Vbound-0}), we find that the most stringent lower bound on the scale
inflation is given by Eq.~(\ref{V*bound}) \footnote{Note that, in this case,
the bound in Eq.~(\ref{Vbound-00}) is not applicable because 
\mbox{$H_{\rm dec}\sim\Gamma_\sigma<H_*$}.}.

Since \mbox{$f\sim 1$}, the only way one can obtain low--scale 
inflation is a very small value of $\varepsilon$. However, 
$\varepsilon$ cannot be arbitrarily small. In fact,
we may obtain a lower bound on $\varepsilon$ as follows:
\begin{equation}
\frac{\delta\sigma_*}{\sigma_*}\leq 1\quad\Rightarrow\quad
\varepsilon\geq\varepsilon_{_{\rm MIN}}\equiv\frac{H_*}{v_0}\,,
\label{epsbound}
\end{equation}
where we have used Eqs.~(\ref{theta}), (\ref{eps}) and (\ref{dsH}) 
and that \mbox{$\sigma_{\rm osc}\lsim v_0$}. 

Another bound on $\varepsilon$ can be obtained with the use of 
Eq.~(\ref{mfbound}), which can be recast as:
\begin{equation}
\varepsilon\geq\varepsilon_{_{\rm MIN}}'\equiv
\frac{\Omega_{\rm dec}}{\pi\zeta\delta^2}\frac{m_\sigma}{m_P}\,.
\label{eps'}
\end{equation}
Comparing the two bounds we find that 
\mbox{$\varepsilon_{_{\rm MIN}}>\varepsilon_{_{\rm MIN}}'$} can be ensured if
\begin{equation}
v_0<(\pi\zeta)\,\Omega_{\rm dec}^{-1}m_P\;,
\end{equation}
which is always satisfied for \mbox{$v_0\leq 10^{14}$ GeV}. For larger values
of $v_0$ the bounds are comparable but only if $m_\sigma$ is not much smaller
than $H_*$. Indeed, it is easy to see that, if \mbox{$m_\sigma<10^{-4}H_*$}
then the bound in Eq.~(\ref{epsbound}) is the tightest for all 
\mbox{$v_0\leq m_P$}. Therefore, in the following we will consider 
$\varepsilon_{_{\rm MIN}}$ as the appropriate lower bound.

From Eqs.~(\ref{V*bound}) and (\ref{epsbound}) and after a little
algebra it is easy to get
\begin{eqnarray}
\,\!V_*^{1/4}\!\geq\!\left(\frac{m_P}{v_0}\right)^2\!\!10^{-13}{\rm GeV}
\!\Rightarrow\!
H_*\!\geq\!\left(\frac{m_P}{v_0}\right)^4\!\!\!10^{-44}{\rm GeV},
\hspace{.3cm} 
\label{Vvbound}
\end{eqnarray}
which means that, in principle, 
{\em the larger $v_0$ is the smaller $V_*^{1/4}$ can become}.
For example, if \mbox{$v_0\sim 10^{10}$ GeV}, then $V_*^{1/4}$ can become as
low as TeV. In practice, however, such low values for the lower bound are 
difficult to attain.

\subsection{Constraining the evolution of the radial field}

We now turn our attention to the evolution of the radial field, which 
determines the value $v(t)$. We will denote the radial field with $\phi$. The
value of $\phi$ during inflation determines $v_*$, while its VEV is $v_0$.

At first it may seem that all that is required is that the radial field $\phi$
has a much smaller value $v_*$, when the cosmological scales exit the horizon 
during inflation, than its value $v_0$ at the onset of the oscillations of the 
curvaton. How $\phi$ changes from $v_*$ to $v_0$ seems not to be important.
However, it turns out that there is indeed an important constraint on the 
behaviour of the radial field, coming from the requirements of the spectral 
index of density perturbations\footnote{We thank C.~Gordon for 
reminding us of this issue.}.

Roughly, the reason is the following. The amplitude of the density 
perturbations is determined by the magnitude of the perturbations of the 
curvaton field, which, in this scenario, apart from the scale of $H_*$ is
also determined by the amplification factor $\varepsilon^{-1}$. The latter 
is determined by the value of the radial field when the curvaton quantum
fluctuations exit the horizon during inflation. A strong variation of the
value of the radial field results in a strong dependence of $\varepsilon(k)$ 
on the comoving momentum scale $k$, which would reflect itself on the 
perturbation spectrum threatening significant departure from scale invariance. 

Let us quantify this.
By definition the spectral index of the density perturbations is
\begin{equation}
n_s-1=\frac{d\ln(\delta\rho/\rho)}{d\ln k}.
\label{ns}
\end{equation}
Now, since \mbox{$\delta\rho/\rho\approx\frac{2}{5}\zeta$} and 
\mbox{$\zeta\propto\zeta_\sigma$} [cf. Eq.~(\ref{zeta})] we find
\begin{equation}
d\ln(\delta\rho/\rho)=d\ln(\delta\sigma_{\rm osc})\,,
\label{Dsosc}
\end{equation}
where we used Eq.~(\ref{zs1}) and considered that $\sigma_{\rm osc}$ does not
have a scale dependence. In view of Eq.~(\ref{dsosc}), Eqs.~(\ref{ns}) and 
(\ref{Dsosc}) show that {\em the contribution of the running of the
amplification factor $\varepsilon^{-1}$ to the spectral index} is
\begin{equation}
n_s-1=\frac{d\ln\varepsilon(N)}{dN}\,,
\label{nseps}
\end{equation}
where $N$ is the number of remaining e-foldings of inflation and we ignored
the other contributions coming from the variation of $H$ and also from the 
curvature of the curvaton potential. 

Using Eqs.~(\ref{theta}) and (\ref{eps}) we see that 
\mbox{$\varepsilon=v_*/v_0$}. Hence, since the observational bound on the
spectral index is \mbox{$|n_s-1|\leq 10^{-2}$}, we find the constraint
\begin{equation}
\frac{d\ln v(N)}{dN}\leq 10^{-2}\quad\Rightarrow\quad
|(\dot\phi/\phi)_*|\ll H_*\;,
\label{nsbound}
\end{equation}
where we used that $v$ is determined by the value of the radial field 
\mbox{$v(t)=\phi(t)$} and that $v_0$ is a constant. From the above we realise
that, in order not to violate the observational constraints regarding the
scale invariance of the density perturbation spectrum, {\em the roll of the 
radial field has to be at most very slow when the cosmological scales exit the 
horizon}. However, the slow roll of $\phi$ cannot remain so indefinitely 
because we need \mbox{$v_0\gg v_*$} to have substantial amplification of the
perturbations (i.e. \mbox{$\varepsilon\ll 1$}) and this has to be achieved in
the remaining \mbox{$N_*\lsim 60$} e-foldings of inflation. Only then can the 
lower bound on the inflationary scale be substantially relaxed. Consequently, 
$\phi$ has to increase dramatically at some point {\em after} the exit of the 
cosmological scales from the horizon. This requirement is crucial for 
model-building. 

\subsection{Concrete examples}

In this section we examine three specific scenarios for attaining low 
scale inflation. In each scenario we study the evolution of the radial field
during inflation, estimate $\varepsilon$ and determine the lowest allowed
inflationary scale, which, according to Eq.~(\ref{H*bound}), is given by
\begin{equation}
H_{\rm min}\sim\varepsilon^{4/5}\times 10^7 \ {\rm GeV}\,.
\label{Hmin}
\end{equation}

There are two ways of obtaining \mbox{$\varepsilon<1$}. One way is to identify
$v$ with the minimum of the potential and assume that $\phi$ has reached it 
before the cosmological scales exit the horizon during inflation. 
This works only if $V(\phi)$ changes with time and, in particular, after the 
cosmological scales exit the horizon. Such a shift of the minimum of $V(\phi)$ 
is realisable if, for example,
the radial field is coupled to some degree of freedom that changes value during
inflation (e.g. the inflaton field itself). 
Another way of obtaining \mbox{$\varepsilon<1$} is to identify 
\mbox{$v(t)=\phi(t)$} and consider that $\phi(t)$ is evolving during inflation,
rolling toward its minimum but not having reached it yet, when the 
cosmological scales exit the horizon. In the following examples we 
employ both methods.

\subsubsection{Symmetry breaking during inflation}

We consider first the case that $\phi$ is initially held at the origin
by an interaction with the inflaton field, being destabilised only when the
inflaton field passes through some critical value\footnote{Another 
possibility is that $\phi$ is locked on top of its potential
hill by being coupled to an oscillating scalar field during inflation 
\cite{lock}. Further, the local maximum can be a point of enhanced 
symmetry for the radial field, which may be originally trapped inside it 
\cite{trap}.}. This situation 
differs from the hybrid inflation mechanism only in that the would-be 
waterfall field $\phi$ is not responsible for the bulk of the inflationary 
potential, but instead is responsible for only a small fraction of it.
The situation was actually described first a few years before hybrid
inflation \cite{kl86}, the  motivation then being the possibility of creating 
cosmic strings on cosmological scales (for a related situation see also 
Ref.~\cite{ars}).
To avoid misunderstanding, we emphasise that we allow  inflation to be either 
hybrid or non-hybrid, it makes no difference to our considerations.

Due to the constraint from the spectral index we need
the roll of the radial field to be slow at first, so that Eq.~(\ref{nsbound}) 
is satisfied when the cosmological scales exit the horizon. However, we need
the roll to accelerate later on so that the final value of $\phi$ can be much 
larger than $\phi_*$. Therefore, we consider an effective running mass model
for the radial field with 
\mbox{$V(\phi)\simeq-\frac{1}{2}m_\phi^2(\phi)\phi^2$}. The effective tachyonic
mass $m_\phi(\phi)$ is such that, when the cosmological scales exit the 
horizon \mbox{$m_\phi(\phi_*)\ll H_*$} and $\phi$ undergoes slow roll. However,
at later times and near the VEV of $\phi$ we require 
\mbox{$m_\phi(v_0)\gg H_*$} so that the total roll of $\phi$ after the exit of 
the cosmological scales is substantial. 

To study this scenario we introduce the following toy-model:
\begin{equation}
V(\phi)=V_0-\frac{1}{4}\lambda\phi^4+\frac{\phi^{n+4}}{m_P^n}\;,
\label{toy1}
\end{equation}
where \mbox{$n\geq 1$}. The above choice of model is somewhat contrived. 
Firstly, in a supersymmetric context 
a negative quartic term in the scalar potential
can be generated only with the aid of additional fields (e.g. see \cite{more}).
Moreover, the quadratic term 
seems to be absent, which implies that supergravity corrections of order 
$\sim\pm H^2\phi^2$ to this potential must be suppressed and also the soft 
mass of the radial field must satisfy the bound
\begin{equation}
m_\phi<\sqrt\lambda H_*\;,
\label{msoft}
\end{equation}
so that, since \mbox{$\phi_*>H_*$} \footnote{The field is considered
to be originally displaced from the top of the potential hill by its quantum 
fluctuations.}, the soft mass term contribution is negligible to the 
potential, when the cosmological scales exit the horizon. Note, however, that, 
if the radial field is kept originally at the origin due to an interaction with
the inflaton, then, immediately after the phase transition which releases the 
radial field, the soft mass term is almost canceled by the interaction term.

Despite the above peculiarities, let us see how such a model performs.
The VEV of $\phi$ is 
\begin{equation}
v_0\sim\lambda^{1/n}m_P\;,
\label{fmin}
\end{equation}
which means that
\begin{equation}
V_0\sim\lambda^{(n+4)/n}m_P^4\;.
\label{V01}
\end{equation}
The effective tachyonic mass of the rolling $\phi$ is
\begin{equation}
m_\phi^2(\phi<v_0)\simeq 3\lambda\phi^2.
\label{mf1}
\end{equation}
The mass of $\phi$ in the vacuum is
\begin{equation}
\tilde m_\phi\sim\lambda^{(n+2)/2n}m_P\;.
\label{mf10}
\end{equation}

Hence, for the above model to work we need that \mbox{$\phi_*\leq\phi_c$}, 
where
\begin{equation}
\phi_c\sim H/\sqrt\lambda\;,
\label{fc}
\end{equation}
is the critical value of the radial field where the field ceases to be light, 
i.e. \mbox{$m_\phi(\phi_c)\sim H$}. Obviously we also need \mbox{$\phi_c<v_0$}.
The field must reach $\phi_c$ before the onset of the curvaton oscillations, 
but not necessarily during inflation. In all cases, though, it is easy to show 
that the slow roll of the field results in
\begin{equation}
\phi_*\sim\phi_c\;,
\label{f*c}
\end{equation}
that is, the field is almost frozen during slow roll. In contrast, after
$\phi_c$ has been reached the field rushes to its VEV in less than an
e-folding (or equivalently a Hubble time). 

Now, in order to avoid inflating the Universe due to the radial field, we 
require 
\begin{equation}
\rho_c>V_0\quad\Rightarrow\quad\phi_c>\sqrt{v_0m_P}\sim\lambda^{2/n}m_P\;,
\label{fcbound}
\end{equation} 
where $\rho_c$ is the density of the Universe when the radial field reaches the
value $\phi_c$ and we used that \mbox{$V(\phi\ll v_0)\sim V_0$}.

Let us now estimate the amplification factor for the curvaton's perturbations. 
The value of $\varepsilon$ is determined by the radial field and is estimated 
as
\begin{equation}
\varepsilon=\frac{\phi_*}{v_0}\sim\frac{\phi_c}{v_0}\,.
\end{equation}
In view of Eqs.~(\ref{fmin}) and (\ref{fcbound}) we find that the minimum 
accessible value is
\begin{equation}
\varepsilon_{\rm min}\sim\lambda^{1/n}.
\label{epsmintoy1}
\end{equation}
Note that, since \mbox{$\phi_c\geq H_*$}, we always have 
\mbox{$\varepsilon_{\rm min}>\varepsilon_{_{\rm MIN}}$} 
[cf.~Eq.~(\ref{epsbound})].

From the above, we see that a low inflationary scale can be accommodated only 
with very small values of $\lambda$. For example, inflation with 
\mbox{$H_*\sim 1$ TeV} can be attained only with \mbox{$\lambda<10^{-5n}$}.
As already discussed when this toy model was introduced, this is only one 
tuning problem of many. 

For soft mass of order 1~TeV the above, in view of Eq.~(\ref{msoft}),
implies immediately that \mbox{$H_*>1$ TeV} always. In fact, using
\mbox{$m_\phi\sim 10^3$ GeV}, Eqs.~(\ref{Hmin}), (\ref{msoft}) and 
(\ref{epsmintoy1}) suggest that
\begin{equation}
H_{\rm min}\sim 10^{\frac{35n+24}{5n+8}} \ {\rm GeV}\,.
\end{equation}
For \mbox{$n=2$} this suggests \mbox{$H_{\rm min}\sim 10^5$ GeV} with
\mbox{$\lambda\sim 10^{-4}$}, whereas for \mbox{$n=4$} we readily obtain
\mbox{$H_{\rm min}\sim 10^6$ GeV} with \mbox{$\lambda\sim 10^{-6}$}.
Note that, for very large $n$ (which, from Eq.~(\ref{fmin}), is equivalent to 
\mbox{$v_0\sim m_P$}), we regain \mbox{$H_{\rm min}\sim 10^7$ GeV}.

Hence, even though somewhat contrived, this toy model shows that the PNGB 
mechanism for amplification of the curvaton's perturbations can indeed in 
principle work in the sense that the lower bound on the inflationary scale
may be relaxed. Of course, in the symmetry breaking case considered 
here, it seems that the radial field must be a flat direction, protected by 
supergravity corrections. One then wonders why is it that the radial field 
itself is not considered as a curvaton candidate, as in Ref.~\cite{laza}.

An interesting variant of the above is considering a model based on the
non-supersymmetric Coleman--Weinberg potential \cite{CW}:
\begin{equation}
V(\phi)=\frac{1}{4}\lambda\phi^4\left(\ln\frac{|\phi|}{v_0}-\frac{1}{4}\right)
+\frac{1}{16}\lambda v_0^4,
\end{equation}
which, for \mbox{$\phi\ll v_0$} corresponds to a negative quartic potential
(this could be thought as the \mbox{$n=0$} case of the above). 

This time we find
\begin{equation}
V_0\equiv V(\phi=0)=\frac{1}{16}\lambda v_0^4.
\label{V01'}
\end{equation}
The effective mass of the rolling $\phi$ is
%
\begin{eqnarray}
m_\phi^2(\phi\ll v_0) & \simeq & -|3+\ln(|\phi|/v_0)|\lambda\phi^2 \nonumber\\
 & \Rightarrow & \tilde m_\phi^2 = 3\lambda v_0^2,
\label{mf1'}
\end{eqnarray}
where $\tilde m_\phi$ is the mass of $\phi$ in the vacuum. 

From Eq.~(\ref{mf1'}) it is evident that the critical value $\phi_c$,
corresponding to the end of slow roll of the radial field, is the same as 
given in Eq.~(\ref{fc}). Then, we readily obtain the constraint
\begin{equation}
\rho_c>V_0\quad\Rightarrow\quad\phi_c>v_0^2/m_P\;.
\label{fcbound'}
\end{equation} 
Note that the above
are equivalent with Eqs.~(\ref{V01}), (\ref{mf1}), (\ref{mf10}) and 
(\ref{fcbound}) with the substitution: 
\mbox{$v_0\rightarrow\lambda^{1/n}m_P$} [cf. Eq.~(\ref{fmin})].

Working in the same manner as previously, we find that
\begin{equation}
\varepsilon\sim\frac{\phi_c}{v_0}\;\;\Rightarrow\;\;
\varepsilon_{\rm min}\sim\frac{v_0}{m_P}\,,
\end{equation}
which, when inserted into Eq.~(\ref{Hmin}) gives
\begin{equation}
H_{\rm min}\sim(v_0/m_P)^{4/5}\times 10^7 \ {\rm GeV}.
\end{equation}
Hence, a relatively small value of $v_0$ can substantially relax the lower 
bound on the inflationary scale. For example, considering that $v_0$ is the
Peccei--Quinn scale \mbox{$v_0\sim 10^{12}$ GeV}, one obtains 
\mbox{$H_{\rm min}\sim 100$ GeV}.

\subsubsection{Smooth curvaton}

We turn our attention now to a more realistic suggestion, which facilitates
the variation of the radial field $\phi$ through a coupling to the inflaton 
field $s$. What we have in mind is a model similar to smooth hybrid inflation
\cite{smooth}, where the $F$-term scalar potential is of the form
\begin{equation}
V_F(\phi)=\left(\mu^2-\frac{\phi^4}{m_P^2}\right)^2+\frac{\phi^6s^2}{m_P^4}\,,
\label{VF}
\end{equation}
where $\mu$ is some suitable mass scale.
In contrast to smooth hybrid inflation we will consider that the vacuum
energy responsible for inflation is not due to the above contribution to
the scalar potential but due to some other $s$-dependent potential $V_s$ 
such that 
\begin{equation}
V_*\simeq V_s(s_*)\gg V_F(\phi_*,s_*)\,.
\label{V*V0}
\end{equation}

The full scalar potential of the radial field must also take supergravity 
corrections into account. Considering supergravity corrections, we have
\begin{equation}
V(\phi)\simeq -m_{\rm eff}^2\phi^2+V_F(\phi)\,,
\label{VH}
\end{equation}
where 
\begin{equation}
m_{\rm eff}\sim\max\{H, m_\phi\}\,,
\label{meff}
\end{equation}
being $m_\phi$ is the soft mass of the radial field 
and where we have assumed a negative effective 
mass-squared because the opposite case would send $\phi$ to zero and render 
the angular field (the curvaton) unphysical.

Suppose, at first, that the contribution to the effective potential due to
supergravity corrections is negligible. This is so only if 
\begin{equation}
m_{\rm eff}<\frac{\mu\phi}{m_P}\,.
\label{mf}
\end{equation}
In this case the scalar 
potential for our radial field $\phi$ is the one shown in Eq.~(\ref{VF}).
Assuming that $\phi$ during inflation attains the value  $v_F$ which minimises 
$V_F$, we have \mbox{$\phi\sim v_F$}, where
\begin{equation}
v_F\sim\left\{
\begin{array}{ll}
\mu m_P/s\;, & \quad s>\sqrt{\mu m_P}\\
 & \\
v_0\sim\sqrt{\mu m_P}\;, & \quad s\leq\sqrt{\mu m_P}
\end{array}\right.\;.
\label{vF}
\end{equation}
Therefore, provided $V_s(s)$ is such that the inflaton field $s$ is decreasing 
with time during inflation and also if \mbox{$s_*\gg\sqrt{\mu m_P}$}, we could
have a gradual increase of $\phi$ due to the roll of $s$, after the 
cosmological scales exit the horizon. If the inflaton slow
rolls then it is easy to see that the radial field will slow roll as well, in 
accordance to the requirement in Eq.~(\ref{nsbound}). However, this is a 
drawback of this scenario because it cancels one of the advantages of inflation
model--building under the curvaton hypothesis; that one can have fast--roll
inflation, which does not require 
a flat direction for the inflaton field \cite{liber,lotfi}. 

Thus, in this case we may achieve an amplification factor $\varepsilon^{-1}$
for the curvature perturbations given by
\begin{equation}
\varepsilon=\frac{v_F(s_*)}{v_0}
\sim\frac{\sqrt{\mu m_P}}{s_*}\ll 1\;.
\end{equation}
The above implies that low--scale inflation may 
indeed be achieved with a suitable value of $\mu$. The minimum allowed
value of $\varepsilon$ is determined by the requirement in Eq.~(\ref{mf}), 
which, when considering that \mbox{$\phi\sim v_F$} as given by Eq.~(\ref{vF}),
suggests
\begin{equation}
\varepsilon_{\rm min}\sim\frac{m_{\rm eff}}{\mu}\sqrt{\frac{m_P}{\mu}}\;.
\label{emin1}
\end{equation}
Inserting the above into Eq.~(\ref{Hmin}) and considering that 
\mbox{$m_{\rm eff}\sim H$} one finds the condition
\begin{equation}
H_{\rm min}\mu^6\sim(10^{10} \ {\rm GeV})^7\;.
\label{condition}
\end{equation}
Hence we see that $H_{\rm min}$ is very sensitive to the value of $\mu$. 
Bounds on the allowed range for $\mu$ are obtained as follows.

Firstly one needs to satisfy the bound in Eq.~(\ref{V*V0}).
Assuming \mbox{$\phi\sim v_F$} during inflation we immediately obtain that
\begin{equation}
V_F(\phi=v_F)\sim\left\{\begin{array}{ll}
\mu^4\;, &\quad s>\sqrt{\mu m_P}\\
 & \\
\mu^3s^2/m_P\;, & \quad s\leq\sqrt{\mu m_P}
\end{array}\right.\;.
\end{equation}
Hence, we see that, \mbox{$(V_F)_*\sim\mu^4$}, which implies the upper bound
\begin{equation}
\mu^4<V_*\sim(H_*m_P)^2\;.
\label{mubound1}
\end{equation}
A lower bound on $\mu$ is obtained from Eq.~(\ref{mf}), when one considers
\mbox{$\phi_*\sim v_F(s_*)$} for \mbox{$s_*>\sqrt{\mu m_P}$}. Thus, one
obtains the bound
\begin{equation}
\mu^3>m_{\rm eff}^2m_P\;.
\label{mubound2}
\end{equation}
Note that, in view of Eq.~(\ref{emin1}), satisfying the above bound guarantees
that \mbox{$\varepsilon_{\rm min}<1$}.

Eqs.~(\ref{mubound1}) and (\ref{mubound2}) result in the following allowed 
range for $\mu$:
\begin{equation}
10.67\leq\log(\mu/ \ {\rm GeV})\leq 11.13\;.
\label{murange}
\end{equation}
Using Eq.~(\ref{condition}), the corresponding range for $H_{\rm min}$ is
\begin{equation}
10^4 \ {\rm GeV}\leq H_{\rm min}\leq 10^7 \ {\rm GeV}\;.
\label{Hminrange1}
\end{equation}

The above result shows that relatively low-scale inflation can be achieved in 
this case with the lowest scale \mbox{$H_{\rm min}$} corresponding to the 
highest value of $\mu$ in the range shown in Eq.~(\ref{murange}). We also found
that we need \mbox{$\mu\approx 10^{11}$ GeV} for this scenario to work, which 
suggests that the VEV of $\phi$ may be comparable to the scale of grand 
unification \mbox{$v_0\sim 10^{15}$ GeV} \footnote{Using Eqs.~(\ref{emin1}) and
(\ref{condition}) it can be easily shown that 
\mbox{$\varepsilon_{\rm min}\sim 10^4(H_{\rm min}/10^{10} {\rm GeV})^{2/3}$}, 
while from Eqs.~(\ref{epsbound}), (\ref{vF}) and (\ref{murange}) one readily
obtains that 
\mbox{$\varepsilon_{_{\rm MIN}}\sim 10^{-4}(H_{\rm min}/10^{10} \ {\rm GeV})$}.
Hence, \mbox{$\varepsilon_{\rm min}\gg\varepsilon_{_{\rm MIN}}$} always.
Note also, that, in view of the above and Eqs.~(\ref{delta}) and (\ref{eps'}) 
we find \mbox{$\varepsilon_{_{\rm MIN}}'\simeq\Omega_{\rm dec}
\varepsilon_{_{\rm MIN}}\leq\varepsilon_{_{\rm MIN}}$}.}.

Here we should point out that there is another constraint that needs to 
be satisfied. We have to ensure that the coupling between the radial field
and the inflaton field does not destabilise the slow roll of the inflaton, 
which is necessary for the spectral index bound in Eq.~(\ref{nsbound}) to be
satisfied, since the variation of $\phi$ is determined by the roll of the 
inflaton. Therefore, we need to ascertain that the mixed term in the scalar 
potential does not contribute to the effective mass of the inflaton enough for
slow roll to be disturbed, i.e.
\begin{equation}
\phi_*^3/m_P^2<H_*\sim m_{\rm eff}\;.
\label{sr}
\end{equation}
Comparing the above with the requirement in Eq.~(\ref{mf}) we see that it is
possible to satisfy the above provided Eq.~(\ref{mubound2}) holds. Then, using
that \mbox{$\phi_*\sim v_F(s_*)\sim\mu m_P/s_*$} we find that the slow roll of 
the inflaton is not disturbed provided \mbox{$s>s_f$}, where
\begin{equation}
s_f\sim\mu\left(\frac{m_P}{m_{\rm eff}}\right)^{1/3}.
\label{sf}
\end{equation}
Note that, by virtue of Eq.~(\ref{mubound2}), \mbox{$\sqrt{\mu m_P}<s_f<m_P$}. 
Therefore, in the range \mbox{$s_f<s\leq m_P$} the inflaton slowly rolls 
toward the origin resulting also in the slow roll of $\phi$ as required. When 
the inflaton reaches the value $s_f$ slow roll inflation is terminated and 
both $s$ and $\phi$ rush toward their VEVs (unless the slow roll is 
interrupted earlier due to the form of $V_s(s)$).

Let us now consider the case when the supergravity corrections are not 
negligible. This case corresponds to:
\begin{equation}
m_{\rm eff}\geq\frac{\mu\phi}{m_P}\,.
\label{mf2}
\end{equation}
Then the scalar potential can be approximated as
\begin{equation}
V(\phi)\simeq 
V_0-m_{\rm eff}^2\phi^2+\frac{\phi^8}{m_P^4}+\frac{\phi^6s^2}{m_P^4}\,,
\label{Vnew}
\end{equation}
where $V_0$ is introduced to avoid 
negative contributions to the energy density, when $\phi$ assumes its VEV.
The minimum of the above potential is given by
\begin{equation}
\,\!v_{\rm eff}\sim\left\{\begin{array}{ll}
\sqrt{m_{\rm eff}/s}\;m_P\;, & \quad s>(m_{\rm eff}m_P^2)^{1/3}\\
 & \\
(m_{\rm eff}/m_P)^{1/3}m_P\;, & \quad s\leq(m_{\rm eff}m_P^2)^{1/3}
\end{array}\right.\;.
\hspace{-1cm}\label{veff}
\end{equation}
This time the VEV of $\phi$ is reached only when $H$ reduced below the soft
mass so that 
\begin{equation}
m_{\rm eff}\rightarrow m_\phi\;\Rightarrow\;
v_0\sim(m_\phi/m_P)^{1/3}m_P\;.
\label{v01}
\end{equation}
For \mbox{$m_\phi\sim 1$ TeV} we have \mbox{$v_0\sim 10^{13}$ GeV}.
It is easy to see that the effective mass of the 
radial field is given by \mbox{$m_{\rm eff}\sim H$}, which means that the field
is driven to the minimum \mbox{$\phi\rightarrow v_{\rm eff}$} in less than an 
e-folding.

From the above we see that, if $V_s(s)$ is such that the inflaton field $s$ is 
decreasing with time during inflation and also if 
\mbox{$s_*\gg(H_*m_P^2)^{1/3}$}, then we could have a gradual increase of 
$\phi$ due to the roll of $s$ after the cosmological scales exit the horizon.
However, when \mbox{$s<(H_*m_P^2)^{1/3}$} the increase of 
\mbox{$\phi\sim v_{\rm eff}$} is halted. In fact, after the end of inflation 
$v_{\rm eff}$ starts to {\em decrease} because 
\mbox{$v_{\rm eff}\propto m_{\rm eff}^{1/3}\sim H^{1/3}(t)$} until 
\mbox{$H\sim m_\phi$} when $\phi$ assumes its VEV $v_0$. This stage of 
decrease of $v$ is counter productive as it increases $\varepsilon$.

This time, considering that \mbox{$\phi\sim v_{\rm eff}$}, we find
\begin{equation}
\varepsilon=\frac{v_{\rm eff}(s_*)}{v_0}\sim
\left(\frac{m_P}{m_\phi}\right)^{1/3}\sqrt{\frac{H_*}{s_*}}\;.
\end{equation}
Since we have \mbox{$s_*\leq m_P$} the above suggests
\begin{equation}
\varepsilon_{\rm min}\sim 10^5\sqrt{H_*/m_P}\;,
\label{emin2}
\end{equation}
where we assumed \mbox{$m_\phi\sim 1$ TeV}. Inserting this into
Eq.~(\ref{Hmin}) it is easy to find that 
\begin{equation}
H_{\rm min}\sim 10^6 \ {\rm GeV}\;.
\end{equation}
Therefore, in this case it is again possible to lower somewhat the scale of 
inflation although this reduction is not dramatic\footnote{From Eqs.~
(\ref{epsbound}) and (\ref{v01}) it is easy to see that 
\mbox{$\varepsilon_{_{\rm MIN}}\sim 10^5(H_{\rm min}/m_P)$}. Comparing this 
with Eq.~(\ref{emin2}) it is straightforward to show that 
\mbox{$\varepsilon_{\rm min}\gg\varepsilon_{_{\rm MIN}}$} always. Moreover,
using Eqs.~(\ref{delta}) and (\ref{eps'}) 
we find \mbox{$\varepsilon_{_{\rm MIN}}'\simeq 10^{-1}\Omega_{\rm dec}
\varepsilon_{_{\rm MIN}}<\varepsilon_{_{\rm MIN}}$}.}. 

Let us discuss briefly a few considerations regarding this case. At first, it
is interesting to obtain the value of $V(v_{\rm eff})$ and compare it 
with $V_*$ as was done in Eq.~(\ref{mubound1}) for the previous case. This 
time, setting $V(v_0)$ to zero, we find
\begin{equation}
V_0\sim\left(\frac{m_\phi}{m_P}\right)^{2/3}
(m_\phi m_P)^2\sim(10^8 \ {\rm GeV})^4
\ll
V_*\;.
\end{equation}
From the above it is evident that the contribution of the radial field to the 
total energy density during inflation is negligible as required.

Regarding the slow roll of the inflaton, using that 
\mbox{$\phi\sim v_{\rm eff}$} it is straightforward to show that
\mbox{$\phi_*^6/m_P^4<m_{\rm eff}^2\sim H_*^2$} during inflation and, 
therefore, the slow roll of the inflaton is not disturbed. However, when
$s$ reduces below $(m_{\rm eff}m_P^2)^{1/3}$ we have 
\mbox{$v_{\rm eff}^6/m_P^4\sim m_{\rm eff}^2\sim H_*^2$} and the slow roll
of both the inflaton and our radial field is terminated. Thus, this time
\begin{equation}
s_f\sim (m_{\rm eff}m_P^2)^{1/3}.
\end{equation}

Finally, lets discuss also the value of $\mu$ in this case. $\mu$ has to 
satisfy the constraint in Eq.~(\ref{mf2}). Using Eq.~(\ref{veff}) and
taking \mbox{$\phi\sim v_{\rm eff}$} it is easy to show that this constraint is
always satisfied if \mbox{$\mu<(H_{\rm min}^2m_P)^{1/3}\sim 10^{10}$ GeV}. For 
a given \mbox{$H_*\geq H_{\rm min}$} one can find a lower bound on $\mu$, above
which the supergravity corrections are always subdominant. 
Indeed, considering also that \mbox{$s_*\leq m_P$}, Eqs.~(\ref{mf2}) and 
(\ref{veff}) suggest that the supergravity corrections cannot dominate for 
\mbox{$\mu>\sqrt{H_*m_P}$}. Since this would also imply that $V(\phi)$ would 
dominate $V_*$ even if the supergravity corrections are negligible, we see 
that \mbox{$\mu>\sqrt{H_*m_P}$} is excluded. The intermediate range of
values for $\mu$ may allow a transition between the two cases during inflation.

To sum up, in the case of the ``smooth curvaton''  we have seen that, for
rather natural values of the parameters, a moderate decrease of the 
inflationary energy scale is indeed possible. However, the model 
suffers from one disadvantage; namely that inflation has to be
of the slow--roll type.

\subsubsection{The curvaton and the waterfall field}

As a final example we consider hybrid inflation \cite{hybrid}, which, apart 
from the inflaton field $s$, introduces another so--called ``waterfall'' field 
$\Phi$, which is responsible for the inflationary vacuum density. During 
inflation $\Phi$ is kept at the origin due to its interaction with the 
inflaton. However, after the inflaton decreases below a critical value, 
$\Phi$ is destabilised; it leaves the origin and rolls rapidly to its VEV $M$. 
At this point inflation is terminated. 

In our example we consider a 
{\em negative} coupling between our radial field and the waterfall field of the
sort that appears in models of inverted hybrid inflation \cite{more}.
This means that we will use the following scalar potential:
\begin{equation}
V(\phi)=V_0-m_{\rm eff}^2\phi^2-g^2\Phi^2\phi^2+\frac{\phi^{n+4}}{m_P^n}\,,
\label{toy3}
\end{equation}
where \mbox{$n\geq 0$}. The minimum of the above potential is located at
\begin{equation}
v=\left[(m_{\rm eff}^2+g^2\Phi^2)m_P^n\right]^{1/(n+2)}.
\label{vPhi}
\end{equation}
Since the effective mass of the radial field is 
\mbox{$m_\phi(\Phi)\sim\sqrt{m_{\rm eff}^2+g^2\Phi^2}\geq m_{\rm eff}\geq H$} 
[cf. Eq.~(\ref{meff})] we expect $\phi$ to roll toward $v$ in less than an 
e-folding. 

During inflation \mbox{$\Phi=0$}, while after inflation \mbox{$\Phi=M$}. 
Therefore, using Eq.~(\ref{vPhi}) it is straightforward to show that
\begin{equation}
\varepsilon=\frac{v_*}{v_0}\sim\left(\frac{m_{\rm eff}}{gM}\right)^{2/(n+2)},
\label{etoy3}
\end{equation}
where we assumed
\begin{equation}
g>m_{\rm eff}/M\,,
\label{gbound1}
\end{equation}
so that \mbox{$\varepsilon<1$}. Using that 
\mbox{$\varepsilon_{_{\rm MIN}}\sim H_{\rm min}/v_0$} 
[cf. Eq.~(\ref{epsbound})], then Eqs.~(\ref{vPhi}) and (\ref{etoy3}) give
\begin{equation}
\frac{\varepsilon}{\varepsilon_{_{\rm MIN}}}\sim
\left[\left(\frac{m_{\rm eff}}{H_{\rm min}}\right)^2
\left(\frac{m_P}{H_{\rm min}}\right)^n\right]^{1/(n+2)}.
\label{epsratio}
\end{equation}

Typically in hybrid inflation the inflationary vacuum density is determined by 
the VEV of the waterfall field $\Phi$. Therefore, we will consider
\begin{equation}
V_*\sim\alpha M^4\quad\Rightarrow\quad H_*\sim\sqrt{\alpha}\,M^2/m_P\;,
\label{alpha}
\end{equation}
where \mbox{$\alpha\leq 1$} is a numerical coefficient. In certain
supersymmetric realisations of hybrid inflation \cite{susyhybrid} $\alpha$ is 
expected to be close but smaller than unity, i.e. \mbox{$\alpha\lsim 1$}.
However, in supernatural \cite{snat} or running--mass hybrid inflation 
\cite{run} the VEV of the waterfall field is typically \mbox{$M\sim m_P$} with 
\mbox{$V_*\sim 10^{10.5}$ GeV}, which suggests that $\alpha$ is extremely 
small; \mbox{$\alpha\sim 10^{-30}$}.

Inserting Eqs.~(\ref{etoy3}) and (\ref{alpha}) 
into Eq.~(\ref{H*bound}) we obtain
\begin{equation}
\left(\frac{M}{m_P}\right)^2>\frac{10^{-11}}{\sqrt{\alpha}}
\left(\frac{m_{\rm eff}}{gM}\right)^{8/5(n+2)}.
\label{Mbound}
\end{equation}

\begin{center}
\begin{figure}
\leavevmode
\hbox{\hspace{-1cm}
\epsfxsize=4.8in
\epsffile{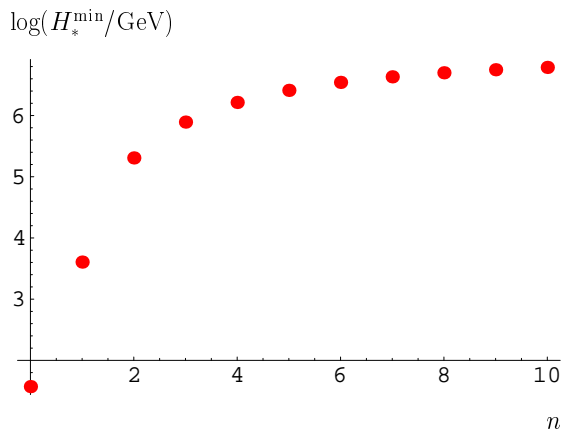}}
\vspace{-9.6cm}
\caption{
Plot of log($H_*^{\rm min}/$ GeV) with respect to $n$ in the case
of a PNGB curvaton whose radial field is coupled to the waterfall field of 
hybrid inflation. It is evident that the larger the $n$ is the tighter the 
lower bound on the inflationary scale becomes. 
\label{Y}}
\end{figure}
\end{center}

Now, during inflation it is easy to show that \mbox{$V(v)\simeq V_0$}, where
\begin{equation}
V_0\sim\tilde m_\phi^2v_0^2
\sim g^4M^4\left(\frac{m_P}{gM}\right)^{2n/(n+2)},
\label{V03}
\end{equation} 
with \mbox{$\tilde m_\phi=m_\phi(\Phi=M)$} being the mass of $\phi$ in the 
vacuum.
To ensure that the density of the radial field is subdominant during inflation
we, therefore, require \mbox{$V_*>V_0$}. Employing Eqs.~(\ref{alpha}) and 
(\ref{V03}) this requirement becomes:
\begin{equation}
g^{n+4}<(\sqrt{\alpha})^{n+2}\left(\frac{M}{m_P}\right)^n.
\label{gbound2}
\end{equation}
Incorporating the above into Eq.~(\ref{Mbound}) and using Eq.~(\ref{alpha}) we
find:
\begin{equation}
\frac{H_*}{m_P}>
\left[10^{-55}
\left(\frac{m_{\rm eff}}{m_P}\right)^{\frac{8}{n+2}}
\right]^{\frac{n+4}{5n+28}}.
\label{Hmeff}
\end{equation}
Remarkably, Eq.~(\ref{Hmeff}) shows that 
{\em the lower bound on the inflationary scale is independent of $\alpha$, 
that is of the VEV of the waterfall field}. 
In view of Eq.~(\ref{meff}), we obtain
\begin{equation}
\,\!
H_{\rm min}\sim 10^7 \ {\rm GeV}\times\left\{\begin{array}{ll}
\!\!10^{-\frac{176}{(5n+6)(n+4)+4n}}\;, & \;\;n>1\\
 & \\
\!\!10^{-\frac{16(2n+19)}{(5n+28)(n+2)}}\;, & \;\;n\leq 1\\
\end{array}\right.,
\label{Hmin3}
\end{equation}
where we have used that \mbox{$H_{\rm min}\leq m_\phi\sim 10^3$ GeV} if 
\mbox{$n\leq\sqrt{52}-6\simeq 1$}. From the above we see that, for 
\mbox{$n=0$}, $H_{\rm min}$ can be as low as \mbox{$H_{\rm min}\sim 10$ GeV}.
Therefore, low scale inflation is indeed attainable. However, the lower bound
on $H_{\rm min}$ increases with $n$ as can be seen in Fig.~\ref{Y}.

In view of the above results, it can be readily checked from
Eq.~(\ref{epsratio}) that \mbox{$\varepsilon\gg\varepsilon_{_{\rm MIN}}$}
always\footnote{One can also show that 
\mbox{$\varepsilon>\varepsilon_{_{\rm MIN}}'$} for all $n$.}.

There is one more condition which we need to verify. Comparing 
Eqs.~(\ref{gbound1}) and (\ref{gbound2}) we find the condition:
\begin{equation}
\left(\frac{m_{\rm eff}}{M}\right)^{n+4}<(\sqrt\alpha)^{n+2}
\left(\frac{M}{m_P}\right)^n.
\label{gcond}
\end{equation}
For \mbox{$H_*>m_\phi$} we have \mbox{$m_{\rm eff}\sim H_*$} and, in view of
Eq.~(\ref{alpha}), the above is equivalent to \mbox{$H_*<m_P$}, which is 
readily satisfied. For \mbox{$H_*\leq m_\phi$}, however, one needs to check the
above condition for a given soft mass $m_\phi$. Using 
\mbox{$m_\phi\sim 10^3$ GeV} we find that the above condition is equivalent to
\begin{equation}
H_*>10^{-15(\frac{n+4}{n+2})}m_P\;.
\end{equation}
It can be checked that the above lower bound is less stringent than the 
ones appearing in Eq.~(\ref{Hmin3}) and, therefore, Eq.~(\ref{gcond}) is always
satisfied, which means that there is always some available parameter space for
$g$. Indeed, from Eqs.~(\ref{gbound1}) and (\ref{gbound2}) is is easy to find 
the range of $g$ for a given $H_*$:
\begin{equation}
\left(\frac{m_{\rm eff}}{H_*}\right)^2\frac{H_*}{m_P}
<\frac{g^2}{\sqrt\alpha}<
\left(\frac{H_*}{m_P}\right)^{n/(n+4)},
\label{grange}
\end{equation}
with the upper bound corresponding to $H_{\rm min}$. For example, for 
\mbox{$H_*\sim 1$ TeV} and \mbox{$\alpha\sim 1$} the allowed range of $g$ is 
\mbox{$10^{-8}<g<0.01-1$}. Note, however, that, for \mbox{$\alpha\ll 1$},
$g$ is bounded from above as \mbox{$g<\alpha^{1/4}$}. Indeed, for 
\mbox{$H_*\sim 1$ TeV} and \mbox{$M\sim m_P$} we have \mbox{$g\lsim 10^{-8}$}
for \mbox{$n=0$}. Since, in this case, the radial field is probably a modulus,
such a weak coupling is expected.

Due to the interaction between the radial field and the waterfall field it is
possible that the mass of the PNGB curvaton can suddenly increase at the 
end of inflation. This may result in \mbox{$f<1$} and one may wonder whether
further relaxation of the lower bound on the inflationary scale is possible.
However, as we have already discussed [cf. Eq.~(\ref{Hbound-0})], the bound on 
$H_*$ cannot be relaxed, even with \mbox{$f\ll 1$}, if \mbox{$\delta\sim 1$} 
as is clearly the case here.

To summarise, we have shown that coupling the radial field to the waterfall 
field of hybrid inflation may allow low scale inflation with 
\mbox{$H_*\sim 10$ GeV} at best. This is quite a satisfactory result and 
corresponds to a rather realistic scenario, since hybrid inflation is well
motivated by particle theory. Moreover, in contrast to the ``smooth curvaton''
example, we do not require slow roll inflation because, in this 
example, {\em the radial field does not roll at all during inflation}. Hence, 
one may have fast--roll hybrid inflation as discussed in Ref.~\cite{mine}. The
only potential difficulty is arranging for the coupling to be negative
(see, however, Ref.~\cite{more}).

\section{The case of a heavy curvaton}

In this section we are going to considered the so called ``heavy curvaton 
scenario''
 where an increment on the curvaton mass, at some moment after the end of inflation
 but before the onset of the curvaton oscillations, leads to a huge decrease
 of the inflationary scale through the attainment of a very small parameter
 $\delta$ [cf. Eq. (\ref{first})]. We will do so by the implementation of a
 second inflationary period following the idea first presented in \cite{matsuda03}.
 We identify this second inflationary period as the thermal inflation one which
 triggers the increment on the curvaton mass when the flaton field, that responsible
 for the generation of the thermal inflation era, rolls down toward the minimum
 of the potential.

\subsection{The thermal inflation model} \label{thermal}
Thermal inflation was introduced as a very nice mechanism to get rid of some
 unwanted relics that the main inflationary
epoch is not able to dilute, without affecting the density perturbations generated
 during ordinary inflation. As
its name suggests, thermal inflation relies on the finite-temperature effects
 on the ``flaton" scalar potential. A
flaton field could be defined as a field with a mass of the order $10^3$ GeV,
 coming from the soft supersymmetric
contributions\footnote{As an example we are going to focus in a value for
 the gravitino mass $m_{3/2}$ of order $m_{3/2}\sim 10^3$ GeV which comes from
 gravity-mediated SUSY breaking.}, and a vacuum expectation value $M$ much bigger
 than $10^3$ GeV \cite{lyth96, lyth95}.
After the period of reheating following
 the main inflationary epoch, the thermal background modifies the flaton potential
 trapping the flaton field at the origin and preventing it to roll-down
toward $M$ \cite{lazarides86,barreiro96}. At this stage
\begin{eqnarray}
\rho = & \hspace{-4mm} V + \rho_T, & \nonumber \\
P = & -V + \frac13 \rho_T, &
\end{eqnarray}
making the condition for thermal inflation, $\rho + 3P < 0$, valid when the
 thermal energy density $\rho_T$ falls below $V_0$,
which corresponds to a temperature of roughly $V_0^{1/4}$. Thermal inflation
 ends when the finite temperature becomes
ineffective at a temperature of order $m_\chi$, so the number of e-folds is

\begin{eqnarray}
N & = & \ln\left(\frac{a_{\rm{end}}}{a_{\rm{start}}}\right) 
=\ln \left(\frac{T_{\rm{start}}}{T_{\rm{end}}}\right)\sim\nonumber\\
& \sim & \ln\left(\frac{V_0^{1/4}}{m_\chi}\right) 
\sim\frac 12\ln\left(\frac{M}{m_\chi}\right).
\end{eqnarray}
Here we have used the fact that, in a flaton potential of the form
\begin{equation}
V = V_0 - (m_\chi^2 - gT^2) \mid \chi \mid^2 + \sum_{n=1}^{\infty} \lambda_n
m_P^{-2n} \mid \chi \mid^{2n+4},
\end{equation}
where the $n$th term dominates:
\begin{eqnarray}
\tilde{m}_\chi^2 = & \hspace{-2.1cm} 2 (n+1) m_\chi^2, & \\
M^{2n+2} m_P^{-2n} = & [2(n+1)(n+2) \lambda_n]^{-1} \tilde{m}_\chi^2, & \\
V_0 = & \hspace{-0.9cm} [2(n+2)]^{-1} \tilde{m}_\chi^2 M^2.&
\end{eqnarray}
It is worthwhile to mention that the potential is stabilized by non-renormalisable
 terms.  Otherwise, the vacuum
expectation value $M$ would not be much bigger than $\tilde{m}_\chi$, spoiling
 the suppression of the unwanted relics.

Guided by the result in \cite{matsuda03} we proceed to implement a second inflationary
 stage into our curvaton
scenario in order to lower the main inflationary energy scale.  If this second
 epoch of inflation is the thermal inflation
one devised in \cite{lyth96,lyth95,lazarides86} we would be solving not only the issue
 of the ordinary inflation
energy scale but also the moduli problem present in the standard cosmology.

In this new scenario the scalar potential would be composed of the usual potential
 terms for the curvaton and the flaton
fields, as well as for a quartic interaction term between the two fields:
\begin{eqnarray}
V(\chi,\sigma) & = &
V_0-(m_\chi^2-gT^2)|\chi|^2+m_\sigma^2|\sigma|^2
+\lambda|\chi|^2|\sigma|^2+\nonumber\\
& + & \sum_{n=1}^\infty\lambda_n m_P^{-2n}|\chi|^{2n+4}.
\end{eqnarray}
The $\lambda$ term is the one which will increment the mass of the curvaton
 field when the flaton acquires its vacuum
expectation value at the end of thermal inflation.

Let's assume that the usual inflation and its corresponding reheating have already
 happened, so that the flaton and
the curvaton fields are embedded into a radiation bath. Therefore, even when
 the minimum of the potential is located at
$\chi = M_\chi(\sigma_*) \neq 0$ and $\sigma = 0$, $\chi$ is trapped at the
 origin because of the finite-temperature effects and $\mid%
\sigma \mid = \sigma_\ast \neq 0$ because $m_\sigma < H < H_\ast$. Thus,
 the value of the scalar potential at this stage
is:
\begin{equation}
V(\chi = 0, \sigma = \sigma_\ast) = V_0 + m_\sigma^2 \sigma_\ast^2,
\end{equation}
with
\begin{eqnarray}
\tilde{m}_\chi^2 =  & \hspace{-0.8cm} 2 (n+1) (m_\chi^2 - \lambda \mid \sigma
 \mid^2), & \\
M_\chi^{2n+2} m_P^{-2n} = & [(n+2) \lambda_n ]^{-1} (m_\chi^2 - \lambda \mid
 \sigma \mid^2), & \\
V_0 = & \hspace{-3mm} [2 (n + 2)]^{-1} (\tilde{m}_\chi^2 M_\chi^2) \mid _{\sigma = 0}. &
\end{eqnarray}

When the thermal energy density falls below $V_0 + m_\sigma^2 \sigma_\ast^2$
 thermal inflation begins. This period
lasts until the temperature is of the order the effective mass of the flaton
 field which is $\tilde{m}_\chi = %
(m_\chi^2 - \lambda \sigma_\ast^2)^{1/2}$. Note that $\lambda\sigma_*^2<m_\chi^2$
 because otherwise there is no thermal inflation. Then, we obtain a first constraint
 on the value of the parameter $\lambda$:
\begin{equation}
\lambda < \frac{m_\chi^2}{\sigma_\ast^2} \sim \frac{10^{-2} \hspace{1mm}
 {\rm GeV}^2}{H_\ast^2 \Omega_{\rm dec}^2}, \label{const_lambda}
\end{equation}
where we have used the Eq. (\ref{sosc}) and focused on $m_\chi \sim 10^3$ GeV
 which comes from
the gravity-mediated SUSY breaking contributions.

When thermal inflation ends the thermal energy density is no longer dominant.
 The Hubble parameter at the end of thermal
inflation is then associated to the energy density coming from the curvaton
 and the flaton fields:
\begin{equation}
H_{\rm{osc}}^2=\frac{\rho_T+V(\chi=0,\sigma=\sigma_\ast)}{3m_P^2} 
\sim\frac{m_\chi^2M^2}{3m_P^2} ,
\end{equation}
so that
\begin{equation}
H_{\rm{osc}} \sim 10^{-16} M, \label{tic}
\end{equation}
and therefore the parameter $f$ [cf. Eq. (\ref{f})] is
\begin{equation}
f \equiv \frac{H_{\rm osc}}{\tilde{m}_\sigma} \sim 10^{-16} \frac{M}{\tilde{m}_\sigma},\label{fagain}
\end{equation}
where $M \equiv M_\chi \mid_{\sigma = 0}$ is somewhere in the range $10^{3}
 \hspace{1mm} {\rm GeV} \ll M \leq 10^{18}%
\hspace{1mm} {\rm GeV}$.

With this so-low value for the Hubble parameter at the end of thermal inflation, the parameter $\delta$ [cf. Eq. (\ref{delta})] is
\begin{equation}
\delta
\sim 10^{-8}\sqrt{\frac{M}{H_\ast}},
\end{equation}
so that the bounds in Eq. (\ref{H*bound}) and Eq. (\ref{Hbound}) become\footnote{In the heavy curvaton mechanism $\varepsilon = 1$ because there is no
amplification of the curvaton perturbations.}:
\begin{eqnarray}
H_\ast & > & 10^{-6}\hspace{1mm}{\rm GeV}\;
\frac{M^{4/5}}{\tilde{m}_\sigma^{4/5}\Omega_{\rm dec}^{2/5}},\label{H_I_1} \\
H_\ast & > & 10^{-7}\hspace{1mm}{\rm GeV}^{1/2}\; 
\frac{M^{3/4}}{\tilde{m}_\sigma^{1/4}\Omega_{\rm dec}^{1/2}}. \label{H_I_2}
\end{eqnarray}
The effective mass of the curvaton
field after the end of thermal inflation, i.e., when $\bar \chi = M_\chi$ and
 $\bar \sigma = 0$ are the average values of the
flaton and the curvaton fields, is
\begin{equation}
\tilde{m}_\sigma = (m_\sigma^2 + \lambda M^2)^{1/2}. \label{cef}
\end{equation}
Note that we are focusing in the case which refers to a final
curvaton decay rate $\Gamma_\sigma$ smaller than the Hubble parameter at the
 beginning of the oscillations $H_{\rm{osc}}$.
This is to make the curvaton field decay before the flaton field so that we
 can keep working in the simplest curvaton scenario where the curvaton field
 oscillates in a radiation background \cite{lyth02,lyth03c}.

Making use of the constraint in Eq. (\ref{const_lambda}) and the expression
 in Eq. (\ref{cef}), and taking into account that the bare curvaton mass $m_\sigma$
 is
smaller than the Hubble parameter $H_{\rm{osc}}$ at the end of thermal inflation,
 we obtain an upper bound on the effective
mass of the curvaton field:
\begin{equation}
\tilde{m}_\sigma < 10^{-1} \hspace{1mm} {\rm GeV} \frac{M}{H_\ast \Omega_{\rm
 dec}}. \label{mboundE1}
\end{equation}
This bound is consistent with Eq. (\ref{H_I_1}).
When the Eq. (\ref{mboundE1}) is applied to the Eq. (\ref{H_I_2}), we obtain
 a lower bound for $H_\ast$ which is consistent with low-energy scale inflation:
\begin{equation}
H_\ast > 10^{-9} \hspace{1mm} {\rm GeV}^{1/3} M^{2/3}. \label{morebounds1}
\end{equation}
The last inequality is stronger than that of Eq. (\ref{H_I_1}) only while the
 effective mass of the curvaton field is
\begin{equation}
\tilde{m}_\sigma > 10^2 \ {\rm GeV}^{10/11} M^{1/11} \Omega_{\rm dec}^{2/11};
 \label{cem}
\end{equation}
otherwise, we still need to consider the expression in Eq. (\ref{H_I_1}).

\subsection{Required parameter space}

Once we have checked the viability of a low-energy scale inflation we proceed
 to investigate the required range of values
for the parameters of the Lagrangian. Hereafter we are going to focus on the
 gravity-mediated SUSY breaking
scheme where the Hubble parameter during inflation is $H_\ast \sim m_{3/2} \sim
 10^3$ GeV. After thermal inflation has ended, the flaton and curvaton fields
 start to oscillate,
eventually decaying into thermalised radiation. The decay process is distinguished
 by the decay rate. The field with the
biggest decay rate will decay first. The flaton and curvaton decay rates are
 given by
\begin{eqnarray}
\Gamma_\chi\approx\gamma_\chi\frac{m_\chi^3}{M^2} & \quad{\rm and}\quad &
\Gamma_\sigma\approx\gamma_\sigma\frac{\tilde{m}_\sigma^3}{m_P^2}, 
\end{eqnarray}
with $\gamma_\chi \lsim 1$ and $\gamma_\sigma \gsim 1$. Since the curvaton mechanism
 must not be modified, the flaton field must
decay well before the curvaton decay. This requires
\begin{equation}
\tilde{m}_\sigma^3 \ll m_\chi^3 \frac{m_P^2}{M^2} \sim \frac{10^{46} \ {\rm
 GeV}^5}{M^2}. \label{mboundA}
\end{equation}
Now, using the expression in Eq. (\ref{H_I_1}), which is valid for $\tilde{m}_\sigma
 \leq 10^2 \ {\rm GeV}^{10/11} M^{1/11} \Omega_{\rm dec}^{2/11}$ [cf. Eq. (\ref{cem})],
 we require
\begin{equation}
\tilde{m}_\sigma > 10^{-11} M \Omega_{\rm dec}^{-1/2}, \label{mboundB}
\end{equation}
in order to obtain low-energy scale inflation. 
Note that, combining the above with Eq.~(\ref{fagain}), we find
\begin{equation}
f<10^{-5}\sqrt{\Omega_{\rm dec}}\ll 1,
\end{equation}
as required by the heavy curvaton scenario.
Similarly to the above, 
using the expression in Eq. (\ref{H_I_2}), which is valid for $\tilde{m}_\sigma
 > 10^2 \ {\rm GeV}^{10/11} M^{1/11} \Omega_{\rm dec}^{2/11}$ [cf. Eq. (\ref{cem})],
 we require
\begin{equation}
\tilde{m}_\sigma > 10^{-40} {\rm GeV}^{-2} M^3 \Omega_{\rm dec}^{-2}. \label{mboundC}
\end{equation}

Thus, for values of $\tilde{m}_\sigma$ less than $10^2 \ {\rm GeV}^{10/11} M^{1/11}
 \Omega_{\rm dec}^{2/11}$ the required range of values for $\tilde{m}_\sigma$
 is:
\begin{equation}
10^{-11} M < \tilde{m}_\sigma < 10^2 \hspace{1mm} {\rm GeV}^{10/11} M^{1/11},
 \label{first_region}
\end{equation}
where the lower bound comes from the Eq. (\ref{mboundB}). The vacuum expectation
 value $M$ is in the range
\begin{equation}
10^{12} \hspace{1mm} {\rm GeV} \lsim M \lsim 10^{14} \ {\rm GeV}.
\end{equation}
where the lower bound comes from the solution to the moduli problem as we will
 see later, and the upper
bound comes from the Eq. (\ref{first_region}).
On the other hand, for values of $\tilde{m}_\sigma$ bigger than $10^2 \ {\rm
 GeV}^{10/11} M^{1/11} \Omega_{\rm dec}^{2/11}$ the required range of values
 for $\tilde{m}_\sigma$ is:
\begin{eqnarray}
\max\{10^2 \hspace{1mm} {\rm GeV}^{10/11} M^{1/11}, 10^{-40} \hspace{1mm}
 {\rm GeV}^{-2} M^3 \} < & & \nonumber\\
<\tilde{m}_\sigma < 10^{15} \hspace{1mm} {\rm
 GeV}^{5/3} / M^{2/3}, & & \label{moreconstraints4}
\end{eqnarray}
where we have used the Eqs. (\ref{mboundA}) and (\ref{mboundC}), and $M$ can
 be up to $m_P$.
We have considered all the other possible constraints on $\tilde{m}_\sigma$
 and found they are irrelevant compared with those in Eq. (\ref{first_region})
 and Eq. (\ref{moreconstraints4}).

\begin{center}
\begin{figure}
\leavevmode
\hbox{\hspace{-2.5cm}%
\epsfxsize=5.5in
\epsffile{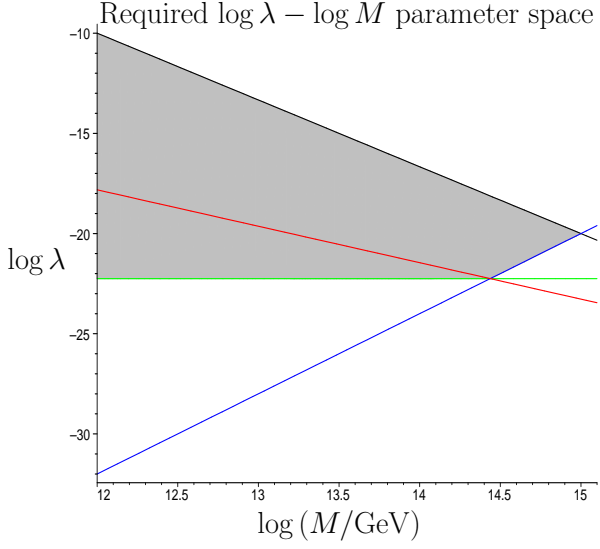}}
\vspace{-11cm}
\caption{
Required $\lambda - M$ parameter space (grey region) as a logarithmic plot.
 The two lines in the middle of the graph correspond to the limits in Eq. (\ref{first_region})
 which are valid up to the meeting point of the three lowest lines. The slanting
 lines correspond to the limits in Eq. (\ref{moreconstraints4}). Note that it
 is impossible to satisfy the conditions in Eq. (\ref{moreconstraints4}) beyond
 the meeting point of the uppest and the lowest lines. \label{lambda_g}}
\end{figure}
\end{center}
\vspace{-.5cm}

Figure \ref{lambda_g} shows the required parameter space $\lambda$ vs $M$ (grey
 region) as a logarithmic plot. We have made use of the definition of the curvaton
 effective mass $\tilde{m}_\sigma$ in terms of the coupling constant $\lambda$
 and the vacuum expectation value $M$:
\begin{equation}
\tilde{m}_\sigma^2 \approx \lambda M^2,
\end{equation}
and the required parameter space $\tilde{m}_\sigma$ vs $M$ studied before. Note
 that for values of $M$ higher than $\sim 10^{15}$ GeV it is impossible to satisfy
 the Eq. (\ref{moreconstraints4}), so our final range for $M$ is
\begin{equation}
10^{12} \hspace{1mm} {\rm GeV} \lsim M \lsim 10^{15}\hspace{1mm} {\rm GeV}.
\end{equation}

The required values for $\lambda$ are in agreement with the upper bound in the
 Eq. (\ref{const_lambda}):
\begin{equation}
\lambda < \frac{10^2 \hspace{1mm} {\rm GeV}^2}{H_\ast^2} \sim 10^{-4}, \label{morebounds2}
\end{equation}
and with the lower bound
\begin{equation}
\lambda > \frac{H_{\rm{osc}}^2}{M^2}%
\approx \frac{m_\chi^2}{3 m_P^2} \sim 10^{-31}, \label{morebounds3}
\end{equation} 
which follows from $\tilde{m}_\sigma > H_{\rm{osc}}$.

Once we have found the required parameter space for $\lambda$ we must do the
 same for the other relevant parameter of the
Lagrangian: the bare mass of the curvaton $m_\sigma$. The first bound we have
 to take into account is
\begin{equation}
m_\sigma < H_{\rm{osc}} \sim 10^{-16} M, \label{morebounds5}
\end{equation}
which is related to the fact that the oscillations of the curvaton around the
 minimum begin due to the sudden increment
in the curvaton mass at the end of thermal inflation. There are other two bounds on $m_\sigma$ we should take into
account. In order to get these bounds, and to understand the lower bound $M
 \gsim 10^{12}$ GeV,
we must study the solution to the moduli problem.

\subsection{Solution to the moduli problem}

Among the unwanted relics that the inflationary epoch is not able to dilute
 are the moduli. Moduli fields are flaton
fields with a vacuum expectation value of order the Planck mass. The decays
 of the flaton and the curvaton fields
increment the entropy, so that the big-bang moduli abundance, defined as that
 produced before thermal inflation and given by
\cite{lyth96}
\begin{equation}
\frac{n_\Phi}{s} \sim \frac{\Phi_0^2}{10 m_P^{3/2} m_\Phi^{1/2}},
\end{equation}
where $\Phi_0$ is the vacuum expectation value of the moduli fields, gets suppressed
 by three factors:
\begin{equation}
\Delta_{PR} \simeq \frac{g_\ast(T_{PR})}{g_\ast(T_C)} \frac{T_{PR}^3}{T_C^3},
\end{equation} 
due to the parametric resonance process following the end of the thermal inflation
 era, where $T_{PR}$ is the temperature just after the
period of preheating and $T_C$ is the temperature at the end of thermal inflation,
\begin{equation}
\Delta_\chi \simeq \frac{4 \beta V_0 / 3 T_\chi}{(2 \pi^2 / 45) g_\ast(T_{PR})
 T_{PR}^3},
\end{equation} 
due to the flaton decay, where $T_\chi$ is the temperature just after the decay\footnote{This is assuming that the flaton has come to dominate the energy
 density just before decaying.}, and $\beta$ is the fraction of the total
energy density left in the flatons by the parametric resonance process ($\beta
 \sim 1$), and
\begin{equation}
\Delta_\sigma \simeq \frac{4 m_\sigma^2 \sigma_{\rm{osc}}^2 / 3 \Omega_{\rm{dec}} T_{\rm dec}}{(2 \pi^2 / 45) g_\ast(T_\chi) T_\chi^3},%
\label{ds}
\end{equation}
due to the curvaton decay, where $T_{\rm dec}$
is the associated reheating temperature which must be bigger than $1$ MeV
not to disturb the nucleosynthesis process\footnote{We have assumed that $\rho_\chi$ does not change appreciably from the time when $T = T_C$ to the time when $T=T_\chi$. This is a good approximation since $\Gamma_\chi\gg\Gamma_\sigma$.}.
 This enhancement in the entropy depends on the temperature just after the flaton decay
\begin{equation}
T_\chi \simeq \frac{10^{13} \hspace{1mm} {\rm GeV}^2}{M} \gamma_\chi^{1/2},
 \label{temp_flat_decay}
\end{equation}
which is obtained by setting $\Gamma_\chi \sim H$ and assuming that the flaton
 decay products thermalise promptly. 
Thus, the abundance of the big-bang moduli after thermal inflation is:
\begin{eqnarray}
\frac{n_\Phi}{s} \sim & 
\frac{\Phi_0^2}{10 m_P^{3/2} m_\Phi^{1/2} \Delta_{PR} \Delta_\chi \Delta_\sigma} \sim%
\frac{\Phi_0^2 T_\chi^4 T_{\rm dec} T_C^3}{10^5 \beta V_0 m_\Phi^{1/2} m_\sigma^2 \Omega_{\rm{dec}} m_P^{3/2} H_\ast^2} & \nonumber%
\\
& \hspace{-0.5cm} \sim 10^{48} \ {\rm GeV}^8 
\,m_\sigma^{-2}M^{-6} 
\gamma_\chi^2 \left(\frac{\Phi_0}{m_P}\right)^2%
\left(\frac{T_{\rm dec}}{1 \hspace{1mm} {\rm MeV}}\right) 
\left(\frac{T_C}{m_\Phi}\right)^3\times & \nonumber\\
 & \hspace{-1cm}\times\left(\frac{m_\Phi}{10^3%
\hspace{1mm}{\rm GeV}}\right)^{1/2}%
\frac{1}{\beta}
\left(\frac{m_\Phi^2 M^2}{V_0}\right) 
\frac{1}{\Omega_{\rm{dec}}}
\left(\frac{10^3 \hspace{1mm} {\rm GeV}}{H_\ast}\right)^2. & \label{abundance}
\end{eqnarray}
The lower bound
\begin{equation}
m_\sigma \gsim \frac{10^{30} \hspace{1mm} {\rm GeV}^4}{M^3}, \label{entropyb1}
\end{equation}
is obtained when taking into account the restriction $n_\Phi / s \lsim 10^{-12}$ coming from nucleosynthesis \cite{ellis92}.

Let's have a look at the thermal inflation moduli abundance defined as that
 produced after thermal inflation:
\begin{eqnarray}
\frac{n_\Phi}{s} \sim  & 
\frac{\Phi_0^2 V_0^2 / 10 m_\Phi^3 m_P^4}{(2 \pi^2 / 45) g_\ast(T_{PR}) T_{PR}^3%
\Delta_\chi \Delta_\sigma} \sim \frac{\Phi_0^2 V_0 T_\chi^4 T_{\rm dec}}{ 10^9
 \beta m_\Phi^3 m_\sigma^2%
\Omega_{\rm{dec}} m_P^4 H_\ast^2} &
\nonumber \\
& \hspace{-.9cm} 
\sim 10^{-6} \hspace{1mm} {\rm GeV}^4 
m_\sigma^{-2}M^{-2} \gamma_\chi^2%
\left(\frac{\Phi_0}{m_P}\right)^2 \left(\frac{T_{\rm dec}}{1 \hspace{1mm} {\rm
 MeV}}%
\right) 
\frac{1}{\beta}
\;\times & \nonumber\\
& \hspace{-1cm} \times
\left(\frac{10^3 \hspace{1mm} {\rm GeV}}{m_\Phi}\right) \left(\frac{V_0}{m_\Phi^2 M^2}\right)%
\frac{1}{\Omega_{\rm{dec}}}
\left(\frac{10^3 \hspace{1mm} {\rm
 GeV}}{H_\ast}\right)^2. & \label{supp_abun_2}
\end{eqnarray}
To suppress the thermal inflation moduli at the required level $n_\Phi / s \lsim 10^{-12}$ we require
\begin{equation}
m_\sigma \gsim \frac{10^3 \hspace{1mm} {\rm GeV}^2}{M}. \label{entropyb2}
\end{equation}

\begin{center}
\begin{figure}
\leavevmode
\hbox{\hspace{-2cm}%
\epsfxsize=5in
\epsffile{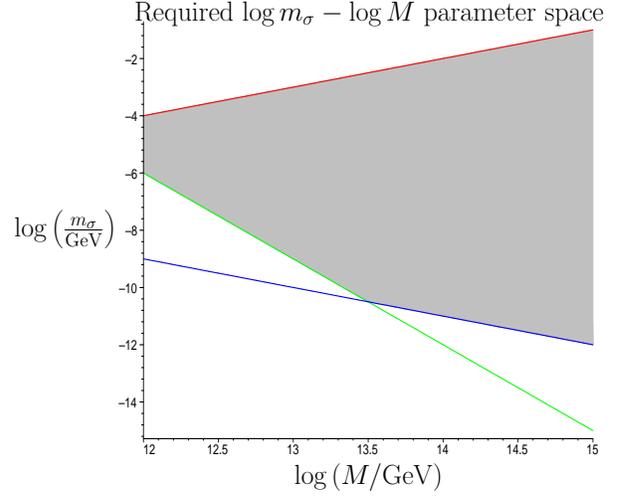}}
\vspace{-10.5cm}
\caption{Required $m_\sigma - M$ parameter space (grey region) from Eq. (\ref{m_space})
 as a logarithmic plot. The smallness of the bare curvaton mass $m_\sigma$ suggests
 the curvaton could be a PNGB. \label{m_g}}
\end{figure}
\end{center}

The Eqs. (\ref{entropyb1}) and (\ref{entropyb2}) are the other two bounds we
 have talked about before, and we have to complement them with that already
 found in Eq. (\ref{morebounds5}):
\begin{equation}
\,\!{\rm max}\left\{\frac{10^{30} \hspace{1mm} {\rm GeV}^4}{M^3}, 
\frac{10^3 \hspace{1mm} {\rm GeV}^2}%
{M}\right\} \lsim m_\sigma \lsim 10^{-16} M.\hspace{-.5cm} \label{m_space}
\end{equation}
This required parameter space is shown in Figure \ref{m_g} (grey region) as
 a logarithmic plot.

\vspace{-7mm}

The Eqs. (\ref{abundance}) and (\ref{supp_abun_2}) give us information about
 the necessary conditions for the suppression of the big-bang and thermal inflation
 moduli, but they are based on the unknown parameters $M$ and
$m_\sigma$. Since we still need to know if the range $M \lsim 10^{15} \hspace{1mm}
 {\rm GeV}$,
required to obtain a low-energy scale inflation, is not forbidden by the requirements
coming from the solution to the moduli problem, we must find a $m_\sigma$-independent
 relation on
$M$. This relation can be found noting that the increment in the entropy due
 to the curvaton
decay (Eq. \ref{ds}) can be written in an alternative way:
\begin{equation}
\Delta_\sigma \simeq \left[\frac{g_\ast (T_{\rm dec})}{g_\ast (T_\chi) (1-\Omega_{\rm{dec}})^3}\right]^{1/4},
 \label{new_Delta}
\end{equation}
so the abundance of big-bang moduli after thermal inflation is:
\begin{eqnarray}
\frac{n_\Phi}{s} \sim 
&
\frac{\Phi_0^2}{10\,m_P^{3/2}m_\Phi^{1/2}\Delta_{PR}\Delta_\chi\Delta_\sigma}
 \sim\frac{10\Phi_0^2T_\chi T_C^3(1-\Omega_{\rm{dec}})^{3/4}}{\beta V_0m_\Phi^{1/2}m_P^{3/2}}
& \nonumber \\
\sim & 
10^{24}{\rm GeV}^3 M^{-3}
\gamma_\chi^{1/2} \left(1-\Omega_{\rm{dec}}\right)^{3/4}
\left(\frac{\Phi_0}{m_P}\right)^2
\times & \nonumber\\
& \hspace{-.7cm}\times
\left(\frac{T_C}{m_\Phi}\right)^3 
\left(\frac{m_\Phi}{10^3\hspace{1mm}{\rm GeV}}\right)^{1/2} 
\frac{1}{\beta}
\left(\frac{m_\Phi^2 M^2}{V_0}\right).
&
\end{eqnarray}
This means that
\begin{equation}
M \gsim 10^{12} \hspace{1mm} {\rm GeV}, \label{lower_M}
\end{equation}
to satisfy $n_\Phi/s \lsim 10^{-12}$. This is the lower bound on $M$ we have
 used throughout the paper.

Of course we might have considered the scenario where there are no moduli fields at all. Without the introduction of the moduli problem the Eqs. (\ref{entropyb1}) and (\ref{entropyb2}) become unnecessary, and the two lowest lines in Figure
 \ref{m_g} disappear. This does not help for the improvement of the required
 range of values for $m_\sigma$ but it does for $\lambda$ as the lower bound
 on $M$ in Eq. (\ref{lower_M}) becomes replaced by $M \gg 10^3$ GeV, which comes from the definition of the flaton fields. In this way the range of values for
 $M$ extends to smaller values well below $10^{12}$ GeV until the coupling constant $\lambda$ eventually reaches the lower bound $10^{-3}$.

The introduction of a period of thermal inflation into our curvaton scenario
 has helped us not only to lower the energy scale of the main inflationary epoch, but also to solve the moduli problem still
 present after ordinary inflation. The
required parameter space has been sketched in Fig. \ref{lambda_g} and \ref{m_g}, and the vacuum expectation value for the flaton field has been showed to be
 in the range $10^{12} \hspace{1mm} {\rm GeV} \lsim M \lsim 10^{15}\hspace{1mm}
 {\rm GeV}$. The required parameter space $\tilde{m}_\sigma - M$ suggests the
 curvaton field could be a PNGB \cite{pngb}. This is because in the presence
 of supergravity all the scalar fields, whose masses are not protected by a
 global symmetry, acquire soft masses of the order of the gravitino mass. The
 smallness of the curvaton mass is in turn because of the very small value for
$H_{\rm{osc}}$. The parameter $H_{\rm{osc}}$ is directly proportional to $M$,
 so the bigger $M$ is, the more possible to obtain a range of values for $m_\sigma$
 compatible with the soft supersymmetric contributions. A higher required value
 for $M$ might be achieved in a scenario with two bouts of thermal inflation
 as suggested in \cite{lyth96}. This would open the possibility of obtaining
 a higher range of required values for $m_\sigma$, but more investigation is
 needed since the scenario proposed in \cite{lyth96} might change with the introduction
 of the curvaton field. We should also look for a mechanism to improve the required
 range of values for the coupling constant $\lambda$ in presence of the moduli
 problem.

\section{Conclusions}

We have presented two different types of curvaton scenario, in which the Hubble
scale of inflation can be much lower than \mbox{$H_*\sim 10^7$GeV}, which 
is the default lower bound for the standard curvaton model \cite{lyth04}. The 
first of these scenarios considers a PNGB curvaton, whose order parameter 
increases after the cosmological scales exit the horizon during inflation. The 
second scenario considers a curvaton, whose mass, being appropriately Higgsed, 
is substantially enlarged at a phase transition after the end of inflation 
(``heavy curvaton''). We have shown that both of these mechanisms are indeed 
able to accommodate inflation scales as low as \mbox{$H_*\sim$ 1~TeV} or even 
lower.

In particular, in the case of a PNGB curvaton, we have derived that the lower
bound on $H_*$ is reduced as $\varepsilon^{4/5}$, where 
\mbox{$\varepsilon\sim v_*/v_0$} is a measure of the growth of the order 
parameter $v$. We have shown that, the rate of variation of the order parameter
(determined by the value of the radial field of our PNGB curvaton) during 
inflation, when the cosmological scales exit the horizon, should not be too 
large because, otherwise, it endangers the scale invariance of the density 
perturbation spectrum. Hence, the radial field must at most slow--roll when the
cosmological scales exit the horizon. However, we need a substantial total
variation of the radial field to achieve an adequately small $\varepsilon$.
This turns up to be a tough requirement to meet in model-building. 
Nevertheless we did come up with a number of successful realisations. 
In fact, we studied three different models for attaining a 
small $\varepsilon$. The first model considers a symmetry breaking during 
inflation, which releases the radial field associated with our PNGB from the 
origin. To preserve scale invariance we have seen that a running tachyonic mass
is required so that our radial field slow rolls at first but runs faster after 
the exit of the cosmological scales from the horizon. Our results showed that, 
by tunning the parameters, a moderate relaxation of the lower bound to the 
inflationary scale can indeed be achieved in the supersymmetric case, when 
the potential is stabilised by non-renormalisable terms. A renormalisable
non-supersymmetric Coleman--Weinberg potential, though, does much better 
without any significant tunning. 
The second model assumes a suitable coupling between the inflaton field 
and the radial field, such that the order parameter of the PNGB is modulated by
the variation of the inflaton. The model is similar to smooth hybrid inflation
in the sense that the radial field, which lies at the temporary minimum of its 
potential, slowly moves away from the origin as its potential changes due to 
the roll of the inflaton. Again we find that a moderate relaxation 
of the lower bound on the inflationary scale is possible. However, one
needs to constrain the vacuum expectation value (VEV) of the radial field 
accordingly, at some appropriate intermediate scale. Another disadvantage is 
that we need slow--roll inflation in order to avoid spoiling the scale 
invariance of the perturbation spectrum, even though these perturbations are 
not due to the inflaton field. Our third model
is the most promising one. Here we introduced a coupling between
the radial field of our PNGB curvaton and the waterfall field of hybrid 
inflation. The latter is kept at the origin during inflation, which means
that our radial field (which again lies at the temporary minimum of its 
potential) also remains constant. At the end of inflation, however, the 
waterfall field rushes toward its VEV. Consequently the potential of our 
radial field changes accordingly and the radial field grows substantially, 
allowing for a really small $\varepsilon$. In this case,
we have found that the we can achieve inflation with Hubble scale as
low as \mbox{$H_*\sim 10$ GeV} regardless of the value of the VEV of the 
waterfall field and only with a reasonable upper bound on the coupling. Note 
also, that the model works well even when considering fast--roll hybrid 
inflation \cite{mine}.

In the heavy curvaton scenario we have worked as follows. We have implemented 
the idea of a thermal inflation epoch, introduced in 
Refs.~\cite{lyth96,lyth95,lazarides86} to solve the moduli problem, as a second
inflationary
period necessary to lower the energy scale of the main inflationary stage.
In our model, a flaton field $\chi$ with bare mass coming from soft 
supersymmetric contributions and vacuum expectation value in the range 
$10^{12} \hspace{1mm}{\rm GeV}\lsim M\lsim 10^{15}\hspace{1mm}{\rm GeV}$, is 
held at the origin of the scalar potential by finite-temperature effects. 
These effects are associated to the thermal background created by the main 
reheating epoch. When temperature falls below $V_0$ thermal inflation begins.
This period of thermal inflation lasts around ten e-folds until the temperature
falls below $m_\chi$ liberating the flaton field to roll away toward the 
minimum of the potential. The curvaton field is coupled to the flaton one so 
its mass is largely increased at the end of thermal inflation. This increment 
is enough to lower the bound on $H_\ast$ to satisfactory levels, without 
sending the non-gaussianity constraint to the limit. However, the energy scale 
of the thermal inflation epoch is very small, requiring in turn a bare mass for
the curvaton field of at most $10^{-4} - 10^{-1}$ GeV. Taking into account the 
soft supersymmetric contributions to $m_\sigma$, the required smallness of 
$m_\sigma$ points toward using a PNGB curvaton to achieve low-scale inflation.
 
Neither of the types of mechanism that we presented is completely compelling.
The first one suffers from the problem of arranging for the appropriate
behaviour of the radial field of the PNGB curvaton. The value of the radial 
field should undergo significant growth in total but, when the cosmological
scales exit the horizon, the field should be at most slowly rolling. One 
possibility is considering a phase transition leading to spontaneous 
symmetry breaking. Just after the transition the effective tachyonic mass of 
the radial field is suppressed and the field is indeed slowly rolling. But
why does the phase transition take place during inflation at a suitable time, 
instead of much earlier or later? That is a good question, though it has not 
prevented several authors from invoking such a phase transition for other 
purposes. It might perhaps be natural if the inflaton and the 
symmetry-breaking field are related in some way. For example, one possibility 
is considering inflation, which lasts only a limited number of e-foldings, 
such as fast--roll inflation \cite{FR} or locked inflation \cite{lock}. In 
this case, it is possible that the required phase transition occurs at (or 
just after) the onset of inflation \cite{KDL}. Other solutions to the problem
of the behaviour of the radial field require particular realisations, which
may well work (as we showed) but the results are highly model dependent. 

The second mechanism suffers from the problem that the mass 
of the curvaton before oscillation, as well as its coupling, have to 
be much smaller than one would expect. In a companion paper \cite{yeinzon04}
it will be shown how this tuning problem can be at least alleviated.

\begin{acknowledgments}
The Lancaster group is supported by
PPARC grants PPA/G/O/2002/00469 and PPA/V/S/2003/00104
and by EU grant  HPRN-CT-2000-00152. DHL is supported by PPARC grants  
 PPA/G/O/2002/00098 and PPA/S/2002/00272.
YR  wants to acknowledge COLCIENCIAS (COLOMBIA) for its full postgraduate scholarship,
 and COLFUTURO (COLOMBIA), UNIVERSITIES UK (UK), and the Department of Physics
 of Lancaster University (UK) for their partial financial support. 
\end{acknowledgments}

\end{document}